\documentclass[prc,twocolumn,secnumroman,tightenlines,amssymb,aps,nobibnotes,
               superscriptaddress,showpacs,balancelastpage,floatfix,nofootinbib]{revtex4}
\usepackage{graphicx}
\usepackage{multirow}           
\usepackage{color}	
\usepackage[utf8]{inputenc}
\expandafter\ifx\csname package@font\endcsname\relax\else
\expandafter\expandafter
\expandafter\usepackage
\expandafter{\csname package@font\endcsname}
\fi
\DeclareRobustCommand\substyle{\name@idx{document substyle}}
\DeclareRobustCommand\classoption{\name@idx{document class option}}
\DeclareRobustCommand\classname{\name@idx{document class}}
\def\name@idx#1#2{{\ttfamily#2}
\index{#2\space#1=\string\ttt{#2}\space#1}\index{#1>#2=\string\ttt{#2}}}

\begin{document}

%\title{Evaluation of the GEF model for fission-fragment yields over an enlarged range}
\title{Benchmark of the GEF model for fission-fragment yields over an enlarged range}

\author{C. Schmitt}
\email{christelle.schmitt@iphc.cnrs.fr}
\affiliation{IPHC CNRS/IN2P3, 23 rue du Loess, B.P.\,28,  F-67037 Strasbourg, France}
\author{K.-H. Schmidt}
\affiliation{CENBG CNRS/IN2P3, Chemin du Solarium, B.P. 120, F-33175 Gradignan, France}
\author{B. Jurado}
\affiliation{CENBG CNRS/IN2P3, Chemin du Solarium, B.P. 120, F-33175 Gradignan, France}

\date{\today}
%\maketitle

\begin{abstract}
\noindent
The GEneral description of Fission observables (GEF) model was developed to produce fission-related nuclear data which are of crucial importance for basic and applied nuclear physics. The investigation of the performance of the GEF code is here extended to a region in fissioning-system mass, charge, excitation energy and angular momentum, as well as to new observables, that could not be benchmarked in detail so far. The work focuses on fragment mass and isotopic distributions, benefiting from recent innovative measurements. The approach reveals a high degree of consistency and provides a very reasonable description of the new data. The physics behind specific discrepancies is discussed, and hints to improve on are given. Comparison of the calculation with experiment permits to highlight the influence of the system intrinsic properties, their interplay, and the importance of experimental aspects, namely instrumental resolution. All together points to the necessity of as selective and accurate as possible experimental data, for proper unfolding of the different influences and robust interpretation of the measurement. The GEF code has become a widely used tool for this purpose.
\end{abstract}
\maketitle

\section{Introduction}

The evolution of a nucleus from a compact configuration into two separated fragments is an intricate puzzle, and it remains a challenge to unambiguously un-fold the influence of the various aspects which enter in play. At the same time, the fission process constitutes a rich laboratory for investigating fundamental nuclear properties  \cite{krappe:2012}, and it is of key importance in applied science, including safeguards, accelerator technology, homeland security, medicine, energy production and waste transmutation at nuclear reactors  \cite{applications:0000}. Fission is also an important process in astrophysical context \cite{goriely:2013}.\\
Modeling fission, in general, implies (i) the definition of the initial conditions as determined by the entrance-channel reaction, (ii) the decay of the fissioning nucleus and its re-arrangement in specific configurations of fragment pairs with corresponding probabilities, (iii) the (prompt) de-excitation of the excited fragments, and (iv) the (slow) decay of the radioactive species towards stability. Models dedicated to fundamental purposes restrict to the first three items, while codes for applications and astrophysics need to cover all aspects. In addition, they need to be computationally fast and flexible, for efficient implementation in general-purpose transport codes modeling the interaction of radiation with nearly everything (see Refs.~\cite{lahet, mcnpx}). Models oriented in the direction of describing the process starting from fundamental principles and quantum mechanics (see Refs.~\cite{schunck:2016, regnier:2016, bulgac:2016, usang:2016}) tremendously developed during the last years. Unfortunately, they cannot afford yet the predictive-power specifications for applied science, and computing time is still prohibitive for large-scale calculations. In parallel, intense work was invested in the development of models and codes suited for use in {\it e.g.} experimental data analysis, data evaluation, applications and astrophysics. They are of various types, going from parameterized systematics \cite{wahl:1988} to more or less phenomelogical or (semi-)empirical models, see Refs.~\cite{madland:1982, lemaire:2005, litaize:2010, lestone:2011, vogt:2011, schmidt:2016, tudora:2017} and therein. The present work investigates the performance of one of these models, the GEneral description of Fission observables (GEF) model \cite{schmidt:2016}, over an extended range of fissioning systems.\\
\\
The GEF code is a semi-empirical model for the description of a (nearly exhaustive) list of fission observables. It combines a set of equations exploiting ideas of quantum mechanics, nuclear dynamics, and statistical mechanics, with a unique set of parameters which were adjusted once to a comprehensive set of experimental data. A particularly rich variety of experimental observations enters the adjustment procedure, including fission excitation functions, fragment properties in mass $A$, charge $Z$, kinetic energy $E_{kin}$, emitted-neutron $\nu$ and $\gamma$-ray multiplicity distributions and energy spectra. This variety is a key point for reaching a consistent description, due to the (sometimes far-reaching) correlations that exist between fission observables. To illustrate the impact of this inter-dependence, we quote an example: The amount and energy of the (prompt and delayed) neutrons are straightforwardly related to the final fragment residues. The latter depend on the so-called pre-neutron fragment ($A$, $Z$) population right after scission, and their excitation energy. The pre-neutron products are determined by, among others, the fragment binding energies, and the generation and sharing of excitation energy at scission. Hence, information about neutron multiplicities and energies can impact the modeling of other observables. A consistent description of as much as possible observables gives confidence in the physics behind a model, as well as in the predictions derived from it. Finally, it helps unraveling hazardeous compensation effects, which could lead to data misinterpretation.\\
\\
A main difference between GEF and most recent codes in the field \cite{lemaire:2005, litaize:2010, vogt:2011, tudora:2017} is that GEF models the fission-fragment production at scission, while other codes rely on the availability of experimental information about the correlation between fragment mass and total kinetic energy $TKE$ \footnote{Some variants of the quoted models can be used with only partial ($A$, $TKE$) information. A parameterization of the $A$ and/or $TKE$ distribution is then implemented \cite{tudora:2017, becker:2013}.}. The present work therefore focuses on fission-fragment yields as calculated with GEF, and evaluates its reliability over a wide range of fissioning system mass, charge, excitation energy $E^*$ and angular momentum $L$. In particular, we exploit recent high-quality experimental measurements, which permit to probe the model in a domain away from the data set used in the adjustment procedure, and in more detail than before. Careful confrontation of  GEF-calculated yields with recent experiments also demonstrates how the code can be used to improve our understanding of the physics behind fission-fragment yields.\\ 
\\
An evaluation of the performance of GEF over an extended range is timely also, as the code is currently used by a widespread community for purposes as various as the treatment of experimental data ({\it e.g.} Ref.~\cite{duke:2016}), guidance for theories ({\it e.g.} Ref. ~\cite{younes:2017}), and interpretation of data ({\it e.g.} Refs.~\cite{khuyagbaatar:2015, pellereau:2017, prasad:2017}). Implementation of GEF in large-scale calculations at Radioactive Beam Facilities and in astrophysics ({\it e.g.} Refs.~\cite{adili:2015, goriely:2015}) has been done also. The GEF code has shown able to examine the consistency of experimental results, and correct - wherever necessary, for incomplete or erroneous measurements \cite{schmidt:2016}. Its predictions on fragment yields can be used as input where no data exists for other models, which are for example focused on the description of fragment de-excitation. The need of supplementing experimental information in such models becomes urgent \cite{jaffke:2017}.\\
\\
In Sec.~\ref{model} a brief description of the main ideas GEF is based on is given, and aspects most relevant for the present concern are discussed. Results of the model calculations are gathered in Sec.~\ref{resul}, and compared with experimental fragment yields measured under various conditions. A critical discussion regarding the achievement by GEF and the level to which it can be probed, on one side, and what we can learn from possible deviation from measurements, on the other side, is proposed. Conclusions are given in Sec.~\ref{concl}.

\section{The GEF model}\label{model}

\subsection{Theoretical framework}

GEF is a semi-empirical model that exploits several general laws of mathematics and physics, combined with empirical information. The philosophy is to bypass the complexity, and computing resources, inherent to a fully microscopic modeling by concentrating
on the essential features, and making use of the regularities observed in experimental data over many fissioning systems and fission quantities. This point of view is based on the high degree of regularity observed in many fission properties (see {\it e.g.} Refs.~\cite{schmidt:2008, bockstiegel:2008}). The formalism is based on solid physics knowledge and well-founded
ideas: (i) the topological properties of a continuous function in multi-dimensional space, (ii) the early localization of nucleonic wave functions in a necked-in shape, (iii) the properties of a quantum oscillator coupled to a heat bath, (iv) an early freeze-out of collective motion, and (v) the application of statistical mechanics. This theoretical frame defines the model in a {\it qualitative} way, and provides the link between different fission observables and the observables of different systems. The parameters entering into the formalism (around 100, of which 50 are most relevant for fission yields) are connected to the underlying physics, and specify the model in a {\it quantitative} way\footnote{Note that popular systematics \cite{wahl:1988} involve a similar number of parameters but for one fissioning system, only.}. These parameters were adjusted once with benchmark experimental data, and are taken the same for all systems. Although GEF does not describe quantities from the microscopic level, it is not a mathematical fit of experimental observations. Also, it preserves the link between different fission observables.\\
In praxis, GEF is a Monte-Carlo code which, starting from a specified, either entrance-channel reaction, or initial compound nucleus, computes the sequential decay of the system, including: possibility of pre-equilibrium emission, pre-scission light-particle evaporation ({\it i.e.} multi-chance fission), ($A$, $Z$) production of the primary (hereafter pre-neutron) excited fission fragments, post-scission evaporation and $\gamma$-ray emission, ($A$, $Z$) population of the secondary  (hereafter post-neutron) products and their radioactive decay wherever suited by delayed neutrons and/or $\gamma$-rays. All along the calculation, the dependence of one step or observable on another one is considered. The user only specifies the reaction or initial compound nucleus of interest.  No parameter is to be adjusted. The results are summarized in a set of output files; an event-by-event list mode data file can be provided upon request. The model is applicable for a wide range of isotopes from $Z$ = 80 to $Z$ = 112 and beyond, from spontaneous fission up to excitation energies of about 100 MeV.

\subsection{Physics behind fragment yields}

A detailed presentation of the modeling of fission in GEF is reviewed in Ref.~\cite{schmidt:2016}. Those features which are most critical for the present focus on fragment yields are discussed below.

\subsubsection{Fission barriers}
A first important  aspect concerns the competition between fission and particle evaporation, which determines the properties of the nuclei that finally go to fission. In the calculation of this competition, the fission barrier plays a central role. In GEF, fission barriers are calculated within the idea of the macro-microscopic approach, and using of the topographic theorem \cite{myers:1996}, which implies that shell corrections at saddle are neglected. The barrier is then given by the sum of the macroscopic barrier (taken from the extended Thomas-Fermi prescription) and the additional binding energy by the empirical ground-state shell correction, plus enhanced pairing correlations at saddle. The approach avoids the uncertainties of theoretical shell-correction energies and allows to discriminate different macroscopic models \cite{kelic:2006}. Note that the reduction of the barrier caused by a finite fissioning system angular momentum is considered \cite{schmidt:2016}.

\subsubsection{Fission channels} 
The second aspect of main importance in this work is the pre-neutron fragment ($A$, $Z$) partition at scission. 
To model this, GEF exploits the basic ideas of the early manifestation of fragment shells \cite{mosel:1971}, the concept 
of quantum oscillators for the fission modes, and dynamical freezing \cite{nifenecker:1980}. In this frame, the fission 
channels are related to the statistical population of quantum oscillators in the mass-asymmetry degree of freedom that 
form the fission valleys. The quantum oscillator of each channel is characterized by three parameters (position, depth, 
and curvature) that are traced back to the macroscopic potential and to shells in the proton and neutron subsystems of both 
fragments, which are assumed to be effective already little beyond the outer saddle \cite{mosel:1971}. These shells are assumed to be the same for all fissioning systems. It is the superposition of different shells and the interaction with the macroscopic potential that create the mass distributions which differ for different systems \cite{schmidt:2008}. Note that these shells also determine the shape (mainly quadrupole deformation) of the nascent fragments at scission. According to Strutinsky-type calculations, the fragment shapes are found to be a linearly increasing function of proton, respectively neutron, number in regions between closed spherical shells  \cite{wilkins:1976}. The influence of fissioning system angular momentum on the parameters of each shell and on the macroscopic potential is taken into account as described in Ref.~\cite{schmidt:2016}. Finally, the charge-polarization (deviation of the $N/Z$ degree of freedom at scission - mean value and fluctuations - from the Unchanged Charge Density value of the fissioning nucleus) is treated by the corresponding quantum oscillator \cite{nifenecker:1980}.\\
Adjustement of the predictions by the above formalism to benchmark experimental mass and charge distributions over a wide 
region of the nuclear chart showed that four fission channels are necessary: the symmetric SL channel, and three asymmetric 
channels ("standard'' S1 and S2, and very asymmetric SA). This empirical adjustement procedure fixed the value of the
parameters of the model related to fragment ($A$, $Z$) yields.

\subsubsection{Energy sorting}
The third aspect which is important for this contribution concerns fission energetics, and in particular the sharing of intrinsic excitation energy between the fragments at scission. It determines the decay of the primary, and population of the secondary, fragments. By the influence of pairing correlations, the nuclear temperature below the critical pairing energy is assumed to be constant \cite{schmidt:2012}. Therefore, the di-nuclear system between saddle and scission consists of two coupled microscopic thermostates \cite{schmidt:2010}. This leads to a sorting process along which the available intrinsic energy and unpaired nucleons before scission are preferentially transferrred to the heavy fragment \cite{schmidt:2011, schmidt:2011a, jurado:2015}. 

\begin{figure*}[!hbt]
\hspace{-1.5cm}
\includegraphics[width=10.cm, height=11.cm,angle=0]{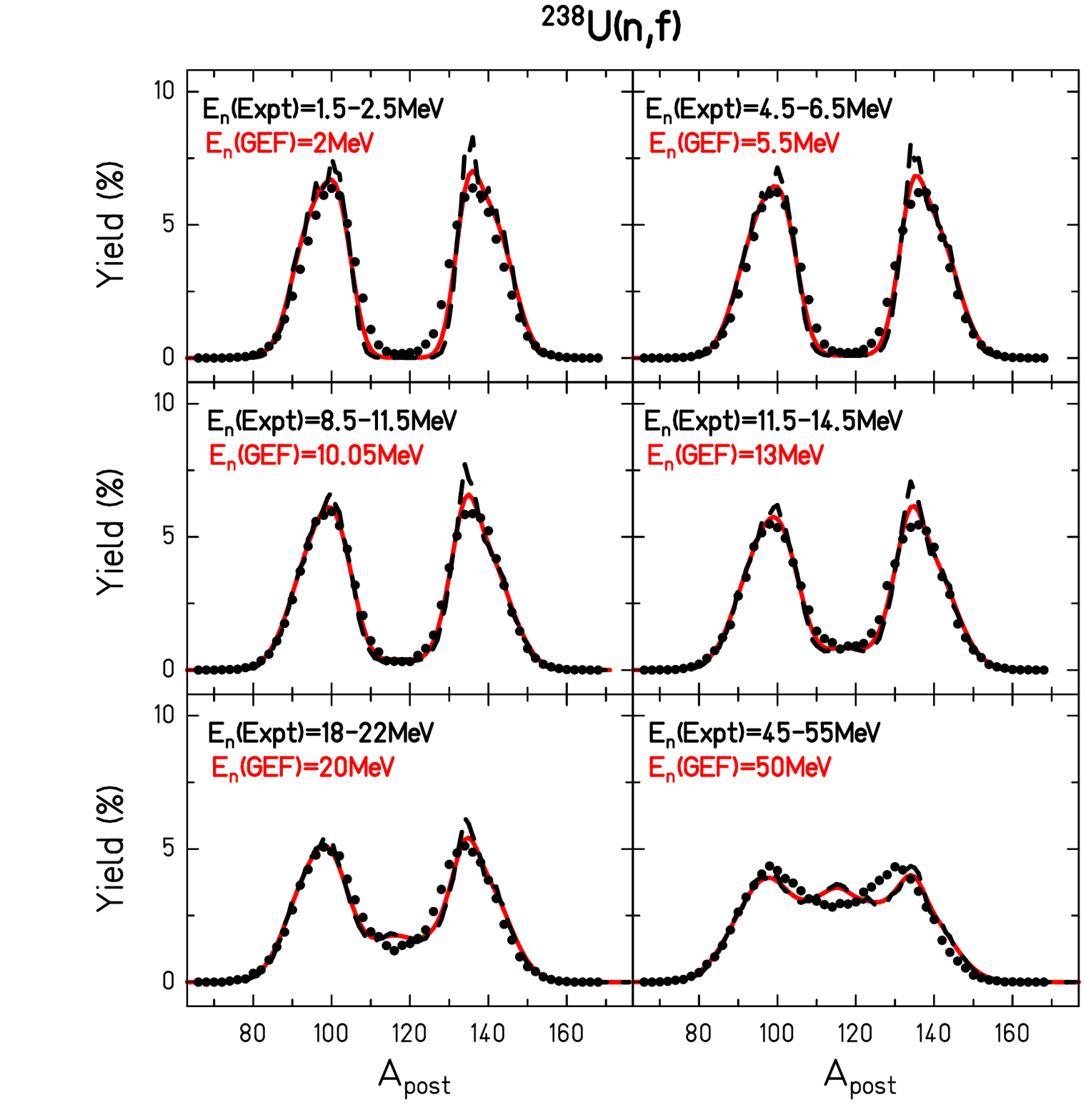}
\hspace{-1.0cm}
\includegraphics[width=10.cm, height=11.cm,angle=0]{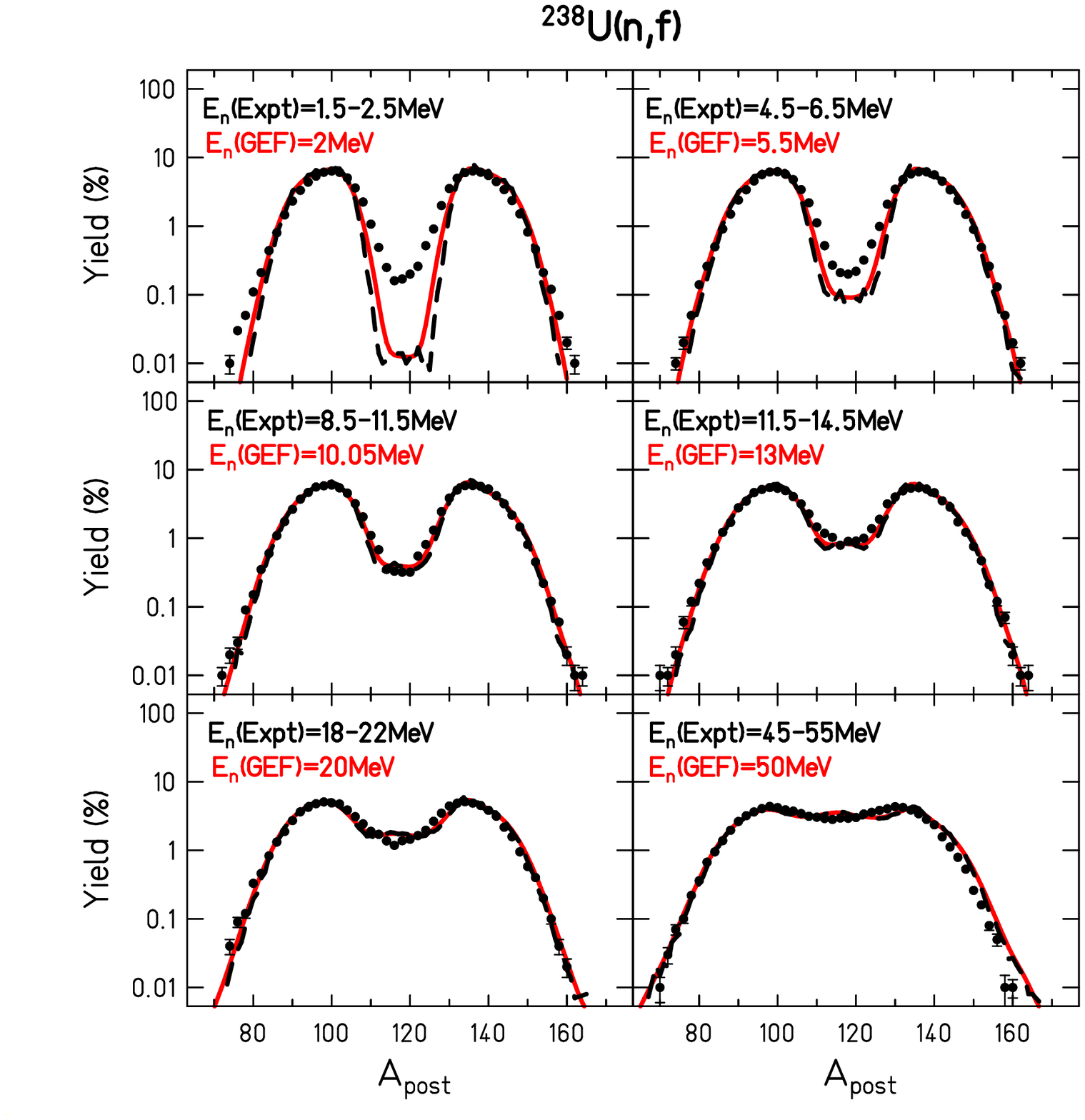}
\vspace{-.20cm}
\caption{(Color online) Fission-fragment post-neutron $A$ distribution for $n$+$^{238}$U at various neutron energies $E_n$ as measured in Ref.~\cite{zoller:1995} (black dots) and compared with GEF calculations without (black dashed line) and with (red full line) account of experimental resolution. The neutron beam energy interval in the experiment as well as the energy used for GEF are indicated. The data are displayed in linear (left) and logarithmic (right) scale.} 
\label{fig1}
\end{figure*}

\section{Experiment vs. GEF}\label{resul}

The performance of the GEF model was studied so far in detail around the region of the data used as benchmark for parameter adjustment. According to the aforementioned high degree of correlation between fission quantities, and the mandatory need of a consistent description, a rich spectrum of observables was considered, including the properties of the outgoing fragments, neutrons and $\gamma$-rays. A large sample of figures is gathered in Ref.~\cite{schmidt:2016}. The study showed the ability of the model in explaining quantitatively many experimental results, over a large range of systems. The present work extends the evaluation into a region of fissioning system mass, charge, excitation energy, and angular momentum, which could, either not been tested so far, or not yet with such a precision. We benefit from innovative, complete, high-quality, measurements. The version of the code used in this work is GEF2016/V1.2.

\subsection{Mass population in fission induced by fast-to-ultra-fast neutrons}\label{zoller}

\begin{figure}[!hbt]
\hspace{-0.2cm}
\includegraphics[width=8.cm, height=5.cm,angle=0]{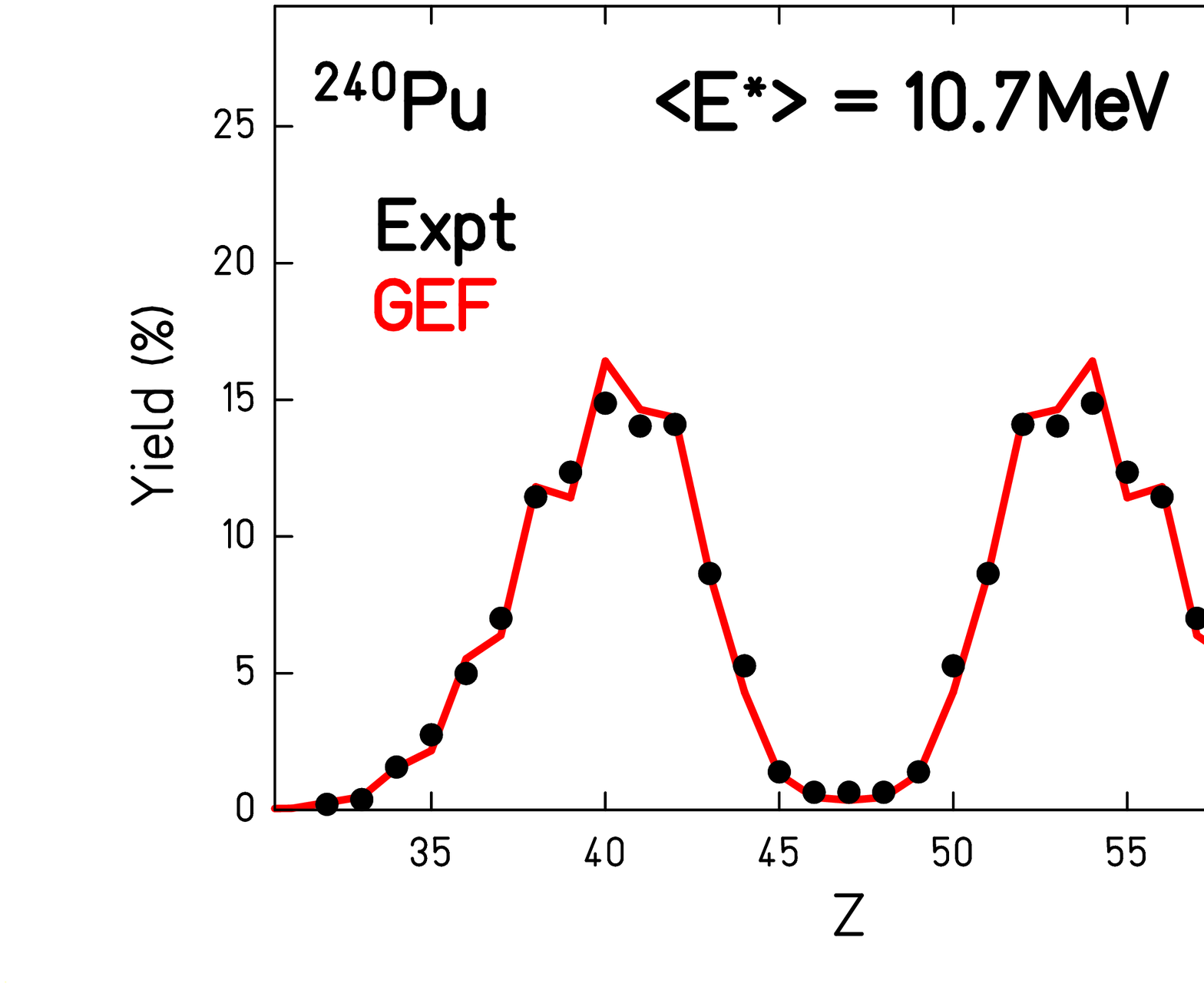}
\includegraphics[width=8.cm, height=5.cm,angle=0]{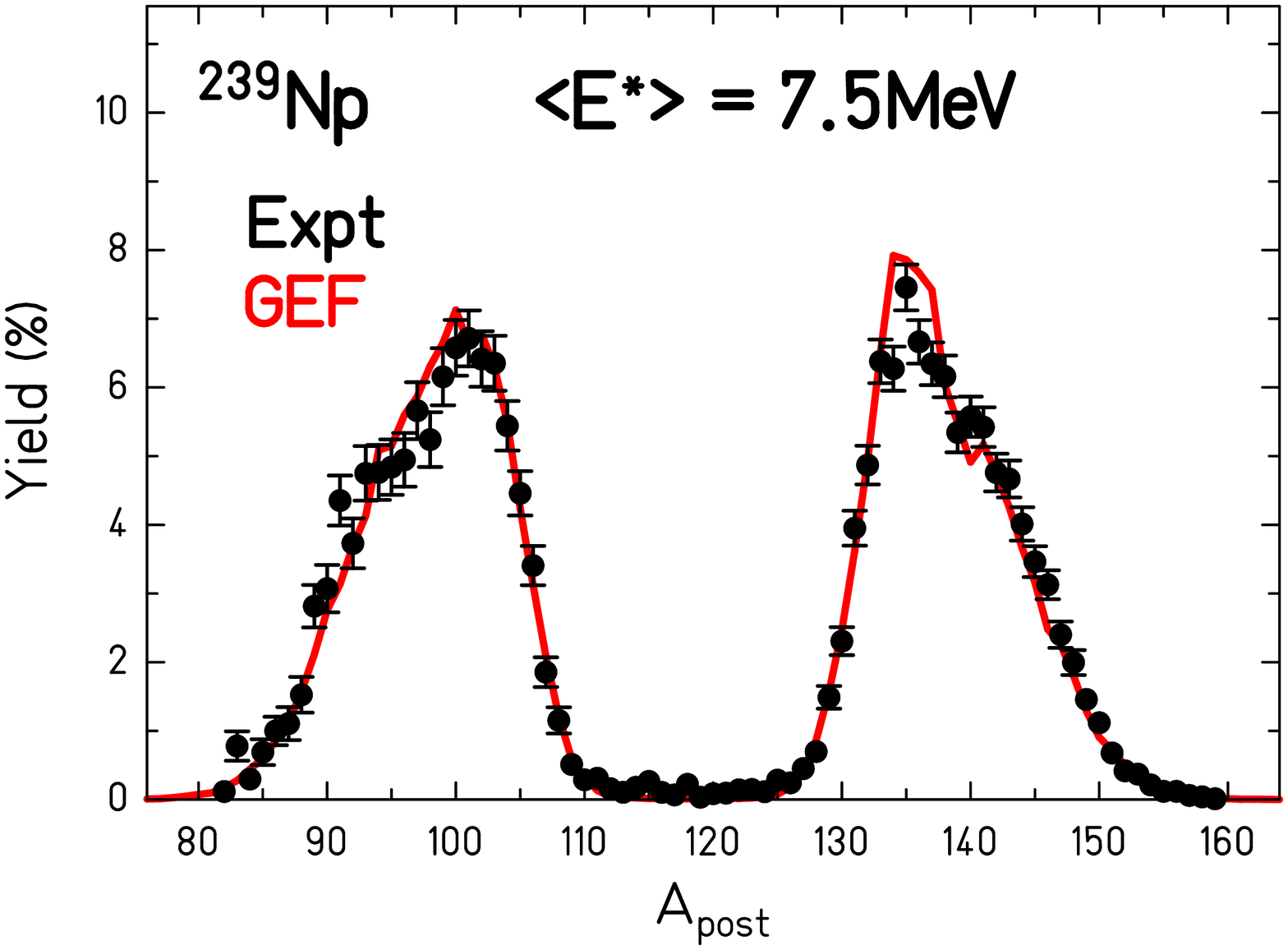}
\vspace{-.40cm}
\caption{(Color online) Fragment $Z$ (top) and  post-neutron $A$ (bottom) distribution as measured at VAMOS (black dots) in low-energy fission of $^{240}$Pu and $^{239}$Np \cite{ramos:2016}, respectively, and compared with GEF calculations (red full line). Whenever not visible, experimental error bars are smaller than the symbols.} 
\label{fig2}
\end{figure}

Before coming to most recent experimental results, we consider the data of Zöller et al. \cite{zoller:1995} on neutron-induced fission of $^{238}$U with ultra-fast neutrons, which permits to extend the regime studied in Ref.~\cite{schmidt:2016}, and are beyond the region used in the model adjustment. Figure~\ref{fig1} compares the measured fragment post-neutron mass $A_{post}$ distributions with GEF calculations, for several neutron energies $E_n$. According to the experimental approach, $E_n$ values are grouped within bins which are 1 to several MeV wide. The calculations are done for a well-defined $E_n$ at the middle of the experimental interval. They are shown without (black dashed lines) and with (red full lines) account of experimental mass resolution ($\sigma$ = 2 assumed). Linear and logarithmic representations are given in order to best appraise the comparison over the entire $A_{post}$ range. Here as well as all along this paper, wherever not visible, error bars are smaller than the symbols. While the overall description is good, including position and width of the asymmetric-fission peaks, and evolution with energy, the calculation (folded with the experimental resolution) shows some deviation from experiment in the symmetric region. Symmetric fission is underpredicted at the lowest energies, and slightly overestimated for ultra-fast neutrons. The finite width of the experimental energy window, and the unknown distribution of $E_n$ within this window, can probably explain part of the discrepancies. We note also that, other experimental methods \cite{crough:1977} yielded results a bit different from those of Zöller et al. \cite{zoller:1995}. However, a deficiency in the modelling in the symmetric region at high excitation energy is clearly observed. Both experimental and theoretical limitations are discussed further in the following sections.\\
The comparison of  Fig.~\ref{fig1} illustrates the importance of experimental resolution. The calculated post-neutron mass distribution exhibits some structures; these are washed once the limited $A_{post}$ resolution is acconted for. Hence, provided that the predicted structures are real, the figure highlights the limited insight that can be get from the data set: To study the physics behind potential structures, improved resolution is mandatory. We note that the data set used in Fig.~\ref{fig1} is representative of the vast majority of the experimental results available in the field till recently. Precise identification could be achieved for selective systems from spontaneous and thermal neutron-induced fission at dedicated separators \cite{rochman:2002}, but only the light fission product was accessible. Radiochemical methods \cite{crough:1977} yield un-ambigous identification, but they depend on radioactive-decay properties, and are thus incomplete.

\subsection{Isotopic production from fission induced in inverse kinematics}\label{isotopic}

\begin{figure*}[!hbt]
\hspace{-1.2cm}
\includegraphics[width=19.cm, height=19.cm,angle=0]{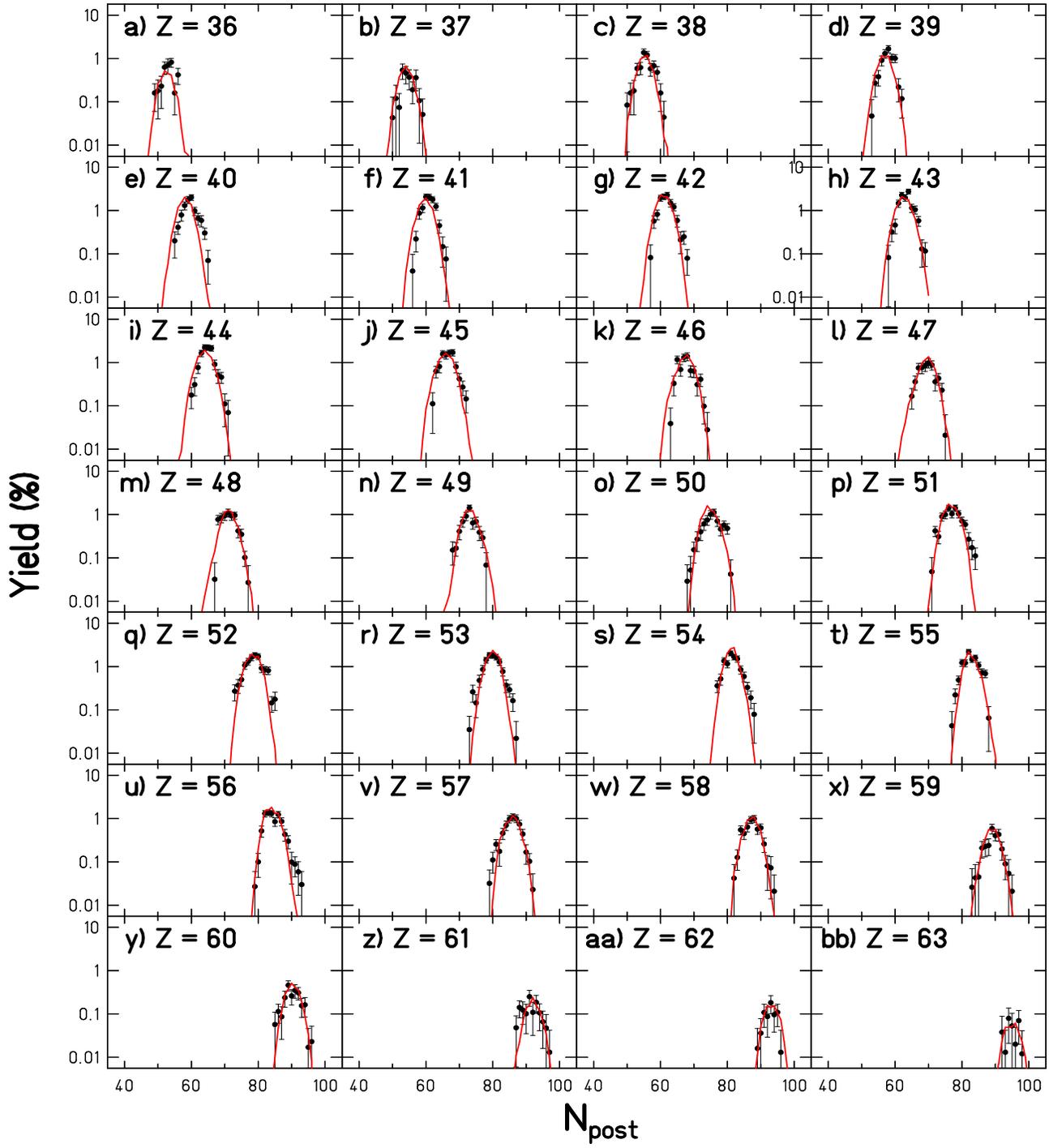}
\vspace{-.10cm}
\caption{(Color online) Fragment post-neutron $N$ distribution for elements between Kr and Eu as measured at VAMOS 
(black dots) in transfer-induced fission of $^{244}$Cm at $<E^*>$ = 23 MeV \cite{ramos:2016}, and compared with the GEF calculation (red full line).} 
\label{fig3}
\end{figure*}

A pioneering experiment performed at GSI \cite{schmidt:2000} demonstrated the substantial gain in data quality when inducing fission in inverse kinematics. The set-up provided unique charge resolution for all (light and heavy) fragments; mass determination was not attempted. This work triggered new-generation experiments, with un-ambigous mass and charge identification over the entire production for fission in the region from pre-actinides to transuranic elements at low-to-medium excitation energies. Results are currently becoming available \cite{caamano:2013, caamano:2015, ramos:2016, pellereau:2017}. Intensive use of these new data is made in this chapter, in order to evaluate the accuracy of GEF predictions and to see what we can learn from potential discrepancies, {\it viz.} which physics is hidden behind.

\begin{figure}[!hbt]
\hspace{-1.3cm}
\includegraphics[width=8.7cm, height=6.2cm,angle=0]{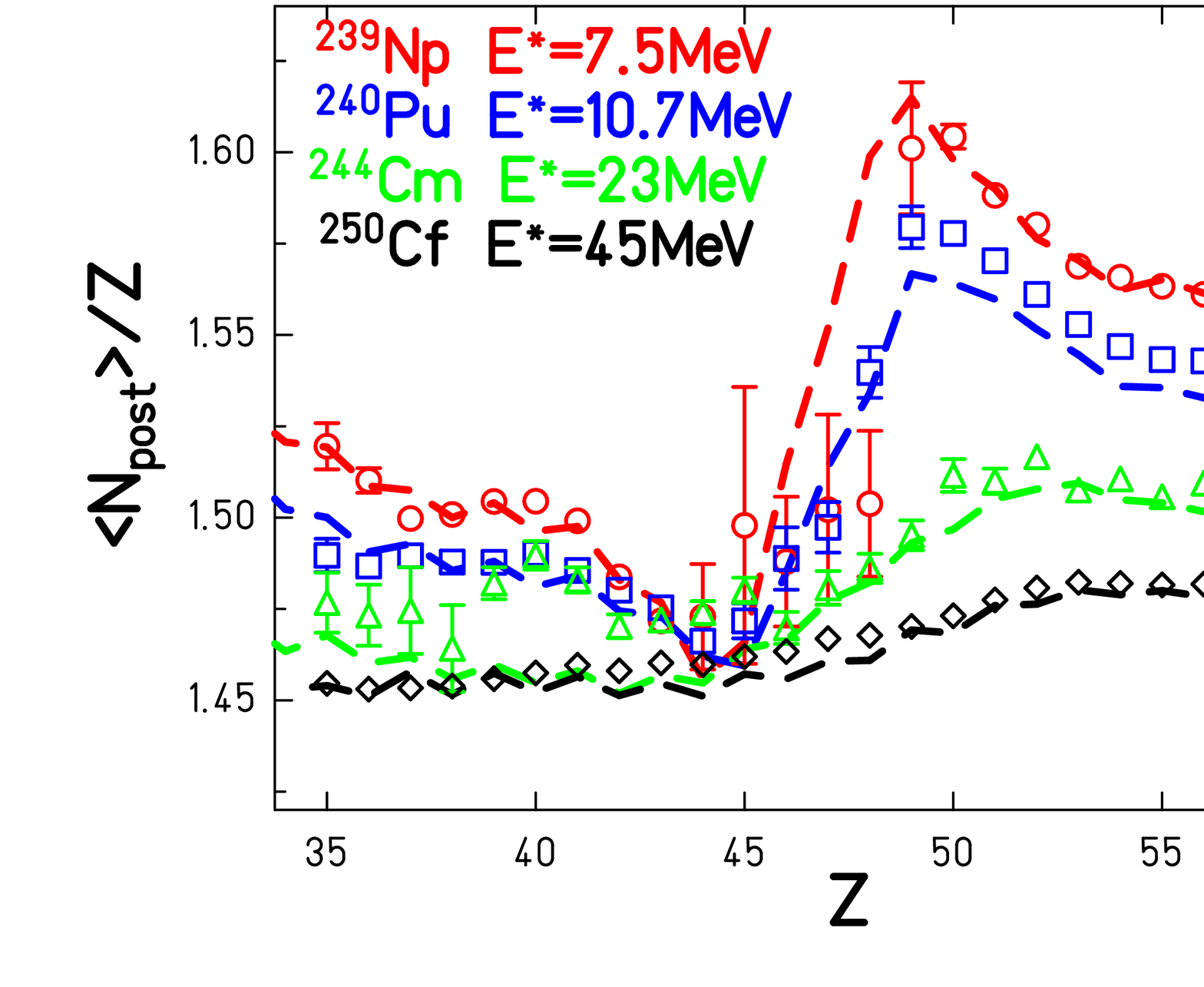}
\vspace{-.70cm}
\caption{(Color online) Mean post-neutron $N$/$Z$ ratio as function of fragment $Z$ for various fissioning systems as measured at VAMOS (symbols) \cite{ramos:2016}, and compared with GEF calculations (dashed lines).} 
\label{fig4}
\end{figure}

\subsubsection{Few-nucleon transfer-induced fission of Pu, Np, Cm and Cf at selective $E^*$}\label{vamos}

The VAMOS campaign at GANIL \cite{caamano:2013, caamano:2015, ramos:2016} investigated fission of transuranic isotopes at low-to-medium excitation energy. Several fissioning systems were populated using few-nucleon transfer reactions from $^{238}$U (6.2MeV/nucleon) + $^{12}$C collisions in inverse kinematics. Fission of the compound nucleus populated in fusion was available concomitantly. The experimental set-up permitted to select the fissioning specie ({\it i.e.} transfer channel) in mass and charge, and determine its excitation energy. Further, the post-neutron mass and charge were uniquely determined for one fragment of the pair using the large-acceptance magnetic spectrometer VAMOS \cite{rejmund:2010}. We consider here the following fissioning systems populated by transfer: $^{239}$Np at $<E^*>$ = 7.5 MeV, $^{240}$Pu at $<E^*>$ = 10.7 MeV, and $^{244}$Cm at $<E^*>$ = 23 MeV, where $<E^*>$ is the mean excitation energy. The spread of the fissioning system excitation energy distribution populated in the transfer reaction varied from $\approx$ 3 to 10 MeV with increasing number of nucleon transfer. Its influence is limited \cite{ramos:2016} and has negligible effect for the concern of this paper. Fission of $^{250}$Cf formed with $E^*$ = 46 MeV by fusion is studied also.

\begin{figure}[!hbt]
\hspace{-1.1cm}
\includegraphics[width=9.0cm, height=6.2cm,angle=0]{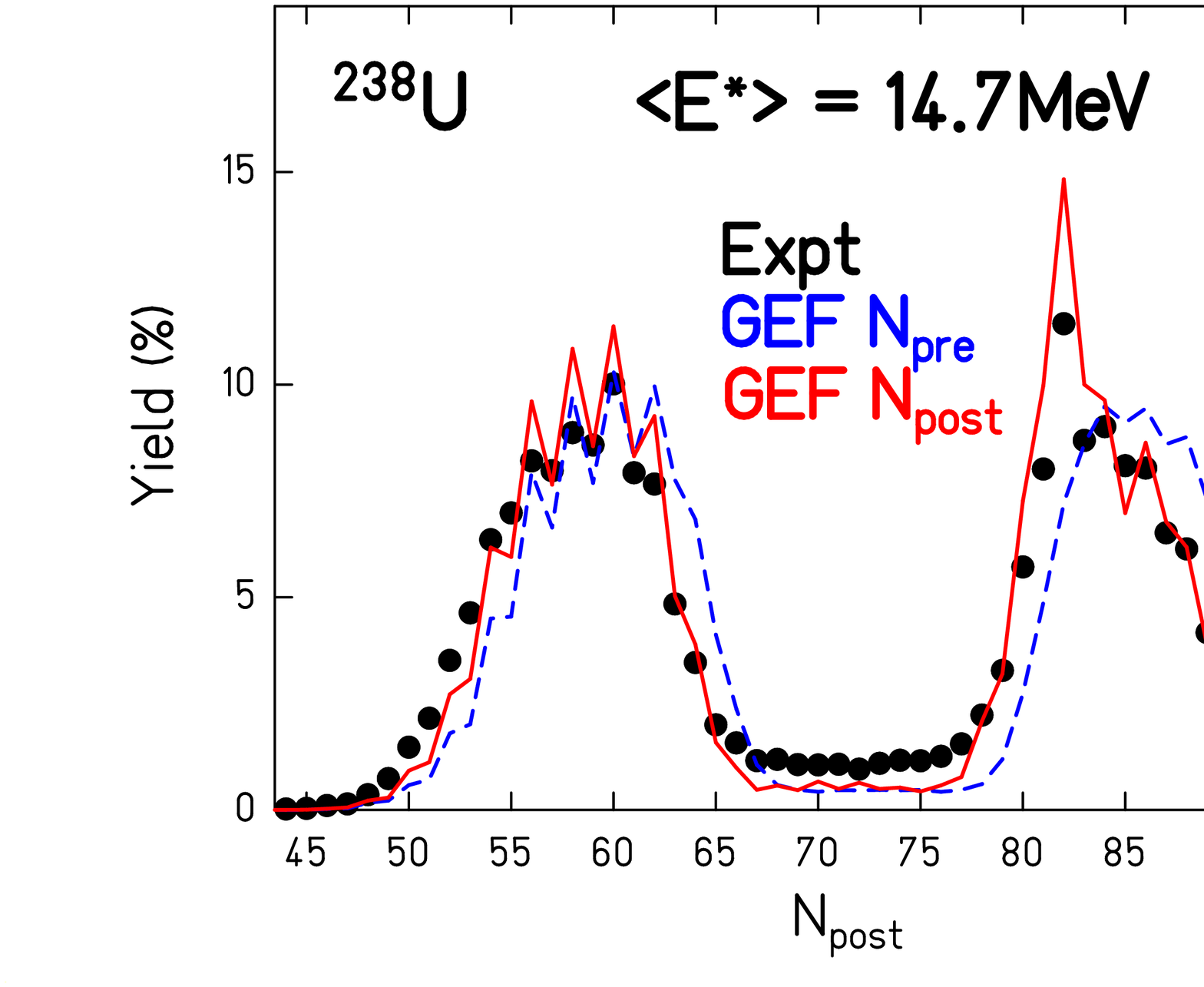}
%\hspace{-1.3cm}
%\includegraphics[width=8.5cm, height=5.5cm,angle=0]{neutron_sofia_U.eps}
\vspace{-.50cm}
\caption{(Color online) Fragment post-neutron $N$ distribution as measured at SOFIA (black dots) in low-energy fission of $^{238}$U \cite{pellereau:2017}, and compared with the GEF calculation for the pre-neutron (blue dashed line) and post-neutron (red full line) $N$ distribution. Experimental error bars are smaller than the symbols.} 
\label{fig5}
\end{figure}

The fragment charge for fission of $^{240}$Pu (top) and post-neutron mass (bottom) distributions for fission of $^{239}$Np are shown in Fig.~\ref{fig2}. The staggering seen in the experimental $Z$ distribution is caused by even-odd effects; it is rather faithfully reproduced by the calculation. As far as post-neutron masses are concerned, the structures predicted by GEF - which could not be probed in Fig.~\ref{fig1}, are clearly visible in the VAMOS data thanks to improved resolution. They are reasonably explained by the model, for both the light and heavy product.\\
The availability of isotopic ($A$, $Z$) information allows an even more stringent test of the model. The calculated post-neutron mass distributions for elements between Kr and Eu for fission of  $^{244}$Cm are compared with the experimental results in Fig.~\ref{fig3}. The agreement is estimated rather satisfactory over the wide range. Apart from possible limitation in the modeling, some deviation, visible at the edges of the distribution for a few elements, was attributed to still not-perfect $Z$ identification in the experiment \cite{farget:2016}.\\
Finally, we consider in Fig.~\ref{fig4} the fragment mean post-neutron $N$/$Z$ ratio as a function of its charge $Z$ for the aforementioned four fissioning systems. This ratio is readily extracted from figures similar to Fig.~\ref{fig3}, corresponding to the mean $N_{post}$ of the distribution for each element. In addtion to the description of $^{244}$Cm which could be anticipated from the previous figure, the calculation fairly describes the evolution from $^{239}$Np to $^{250}$Cf, from about 7 to 50 MeV of excitation energy. We note some deviation at $Z \approx$ 45-48 for $^{239}$Np. The accuracy of the experimental points in the symmetric region where the yields are very small (see Fig.~\ref{fig2}, bottom) may be questioned, before model calculations are revisited. Finally, we refer to Ref.~\cite{caamano:2015, ramos:2016} where still other quantities derived from fragment yields of the GANIL campaign are investigated and compared with GEF. 

\begin{figure*}[!hbt]
\includegraphics[width=18.cm, height=4.cm,angle=0]{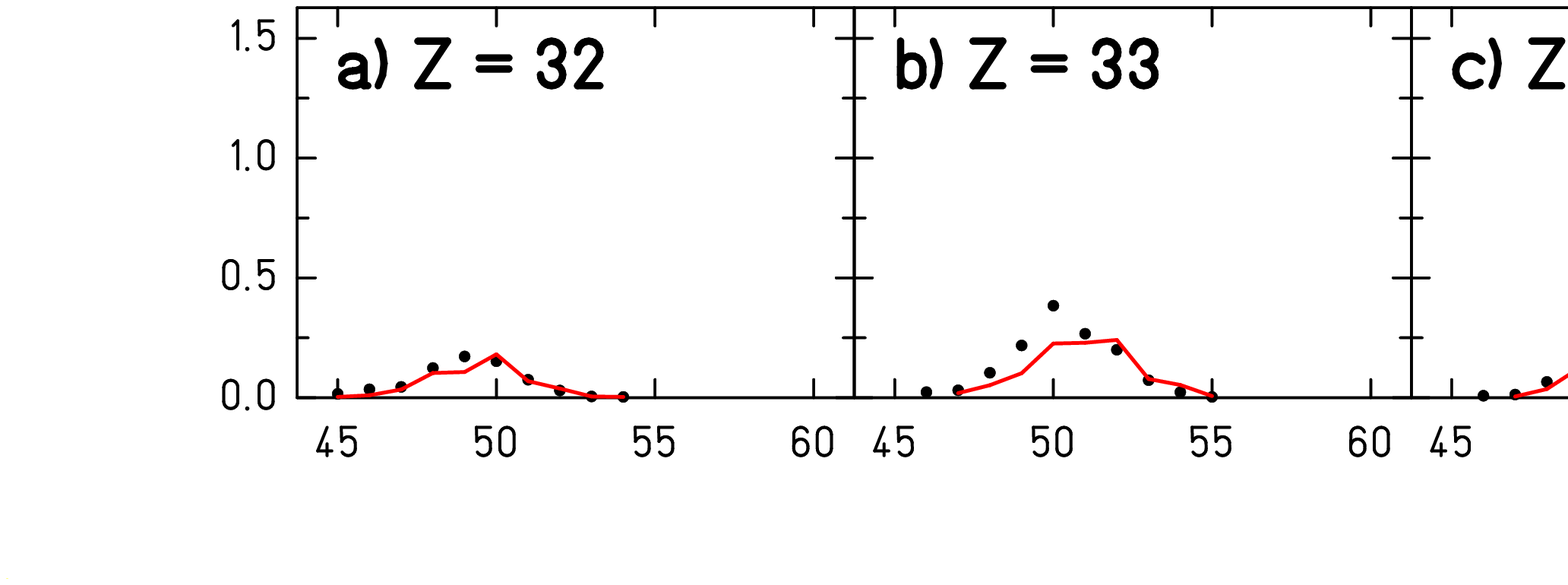}
\vspace{-1.cm}
\includegraphics[width=18.cm, height=4.cm,angle=0]{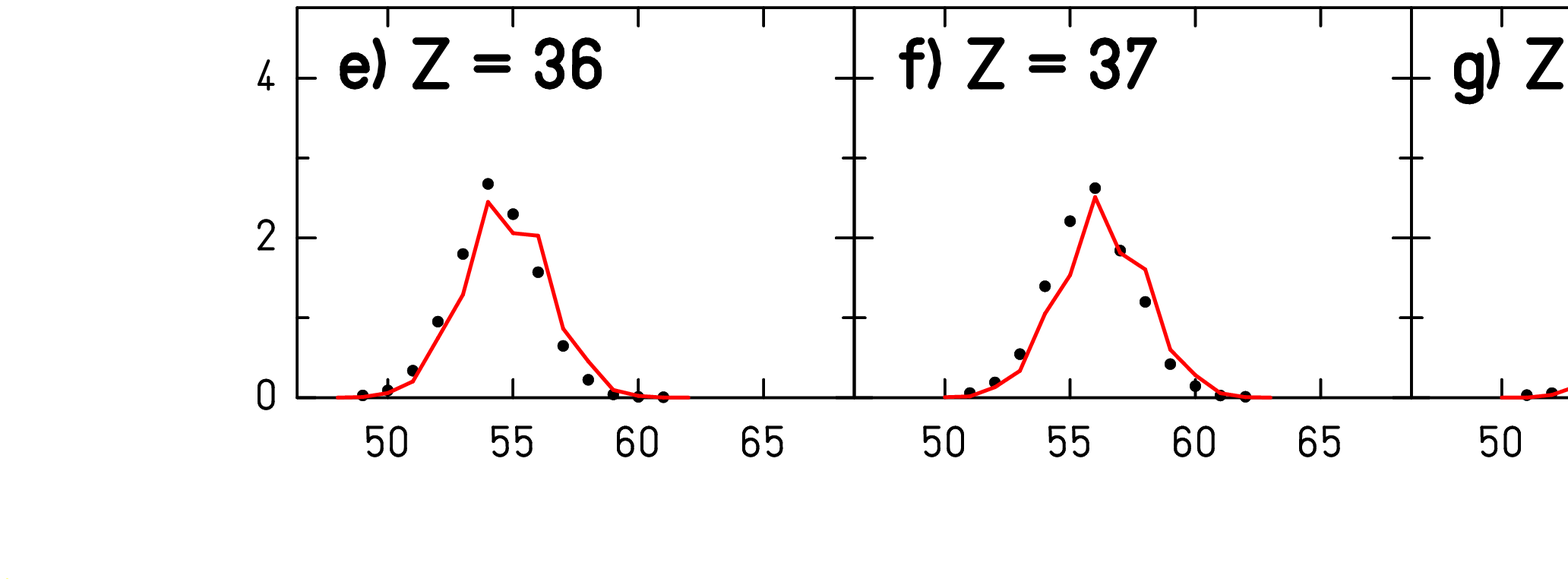}
\vspace{-.70cm}
\includegraphics[width=18.cm, height=4.cm,angle=0]{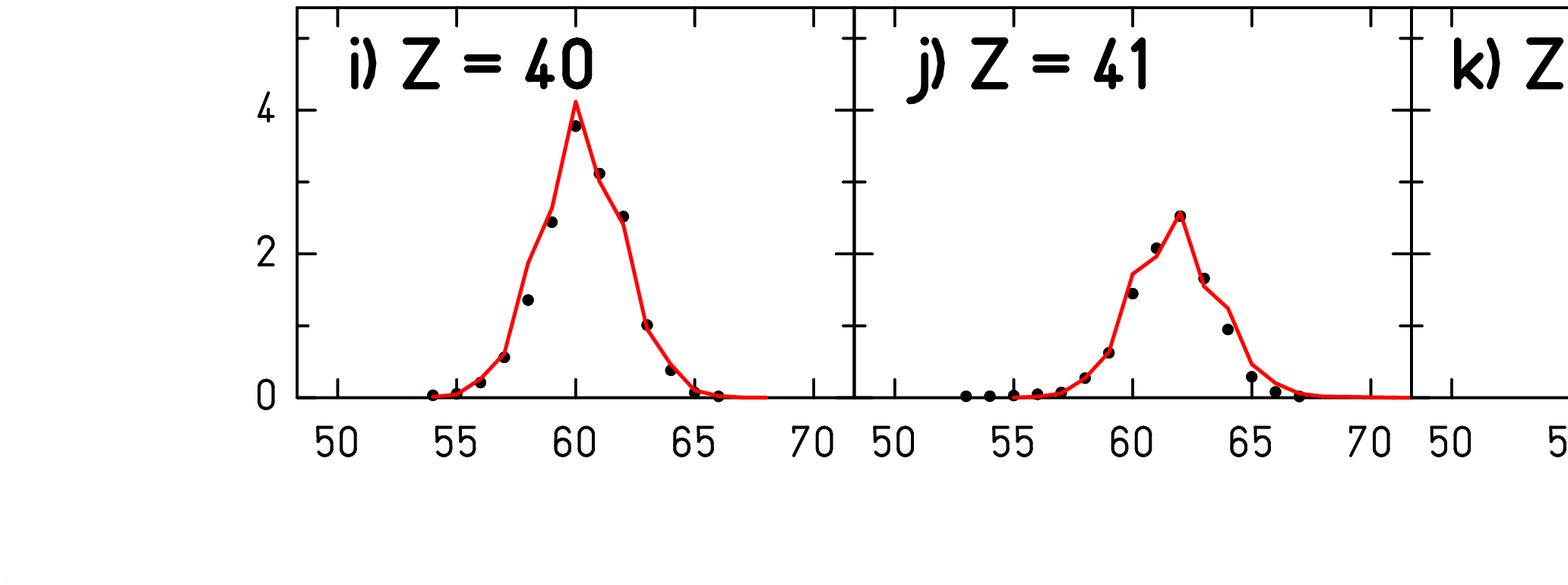}
\vspace{-.70cm}
\includegraphics[width=18.cm, height=4.cm,angle=0]{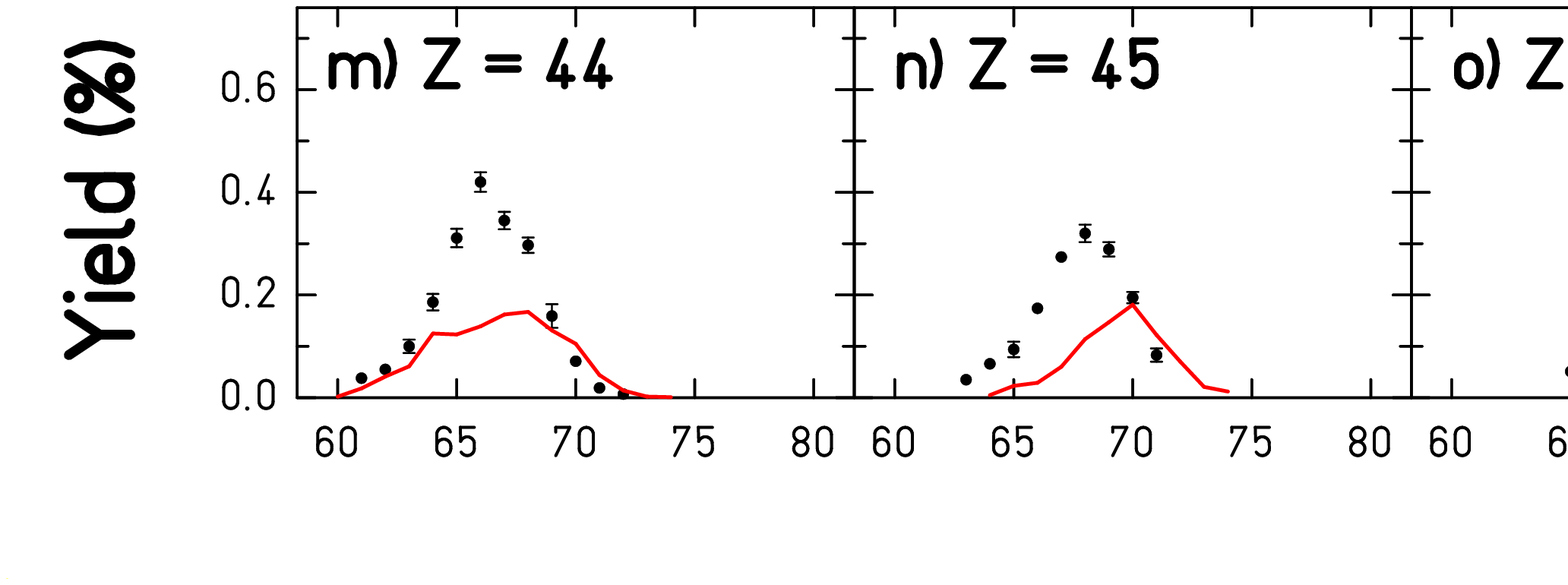}
\vspace{-.70cm}
\includegraphics[width=18.cm, height=4.cm,angle=0]{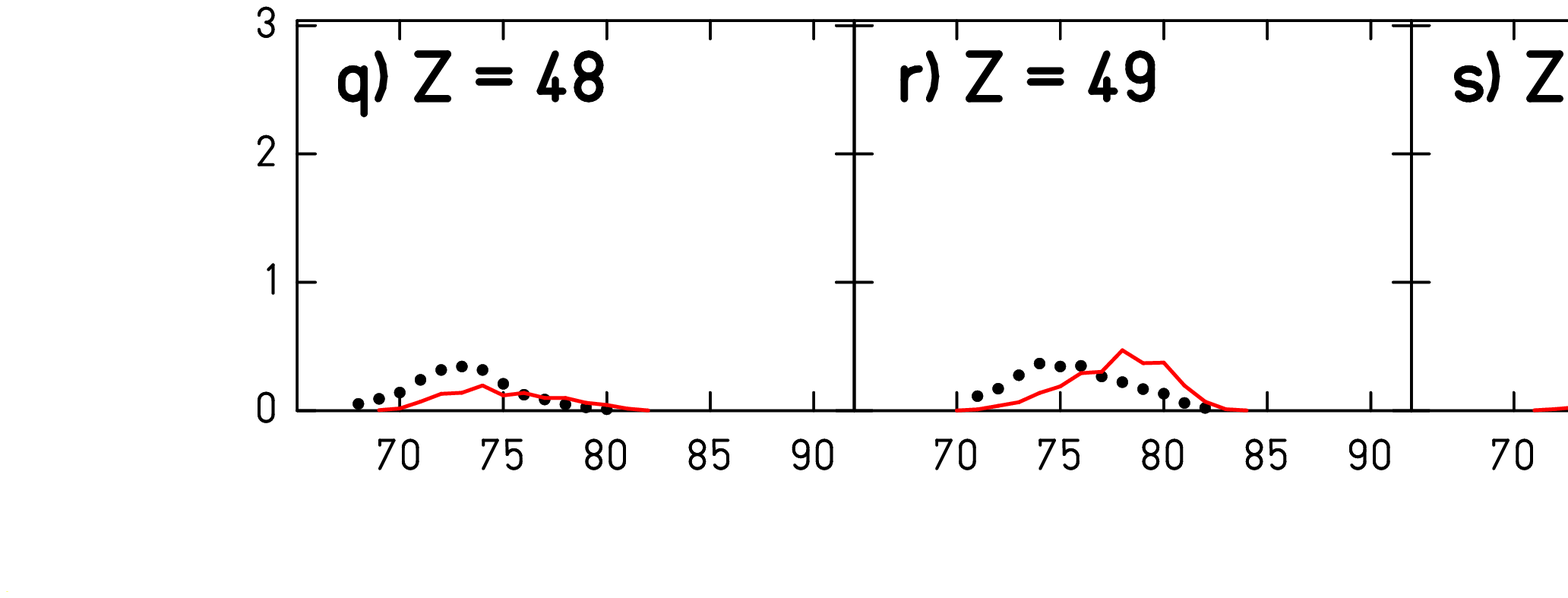}
\vspace{-.70cm}
\includegraphics[width=18.cm, height=4.cm,angle=0]{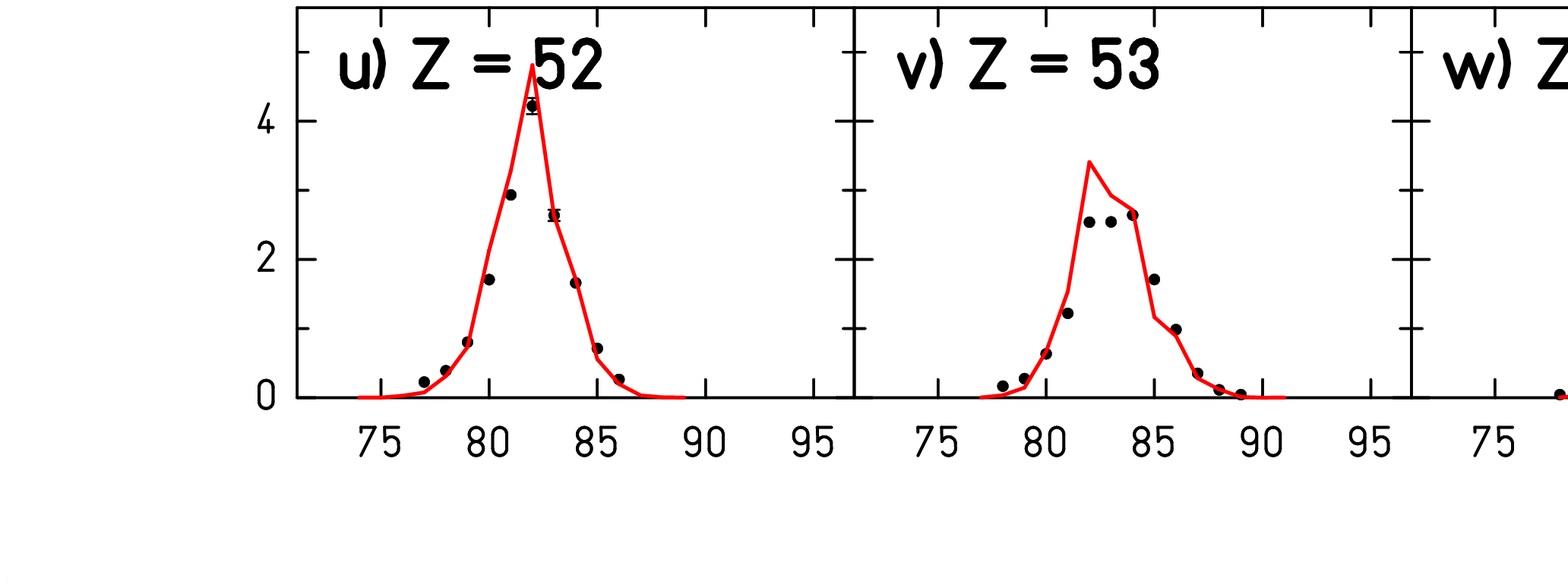}
\vspace{-.70cm}
\includegraphics[width=18.cm, height=4.cm,angle=0]{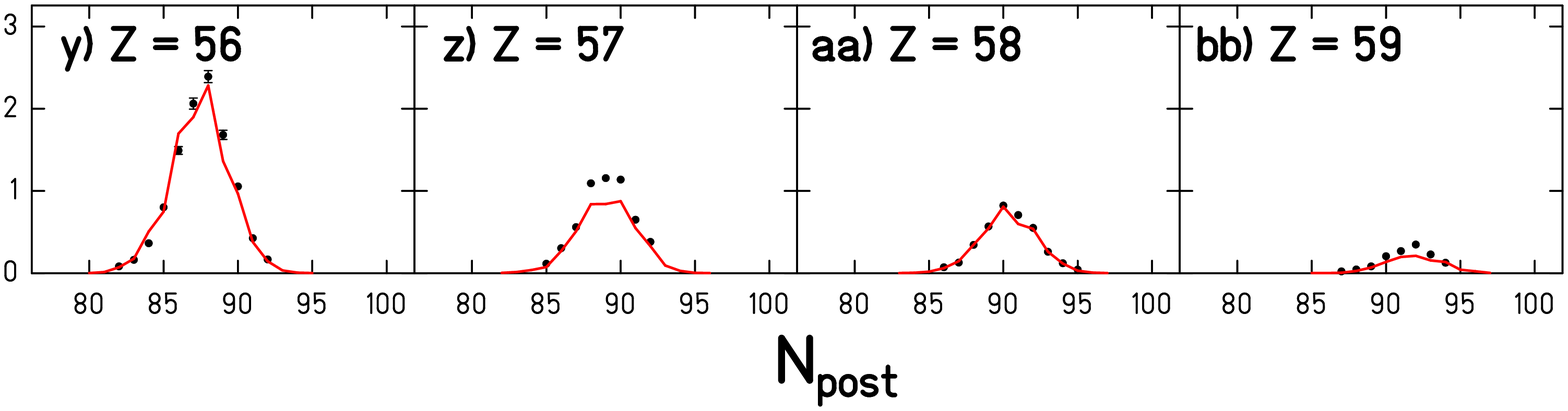}
\vspace{.20cm}
\caption{(Color online) Fragment post-neutron $N$ distribution for elements between Ge and Pr as measured at SOFIA (black dots) in low-energy fission of $^{238}$U \cite{pellereau:2017}, and compared with the GEF calculation (red full line). In most cases, experimental error bars are smaller than the symbols.} 
\label{fig6}
\end{figure*}

\subsubsection{Electromagnetic-induced fission of $^{238}$U}\label{sofia}

\begin{figure}[!hbt]
\hspace{-0.8cm}
\includegraphics[width=9.0cm, height=7.cm,angle=0]{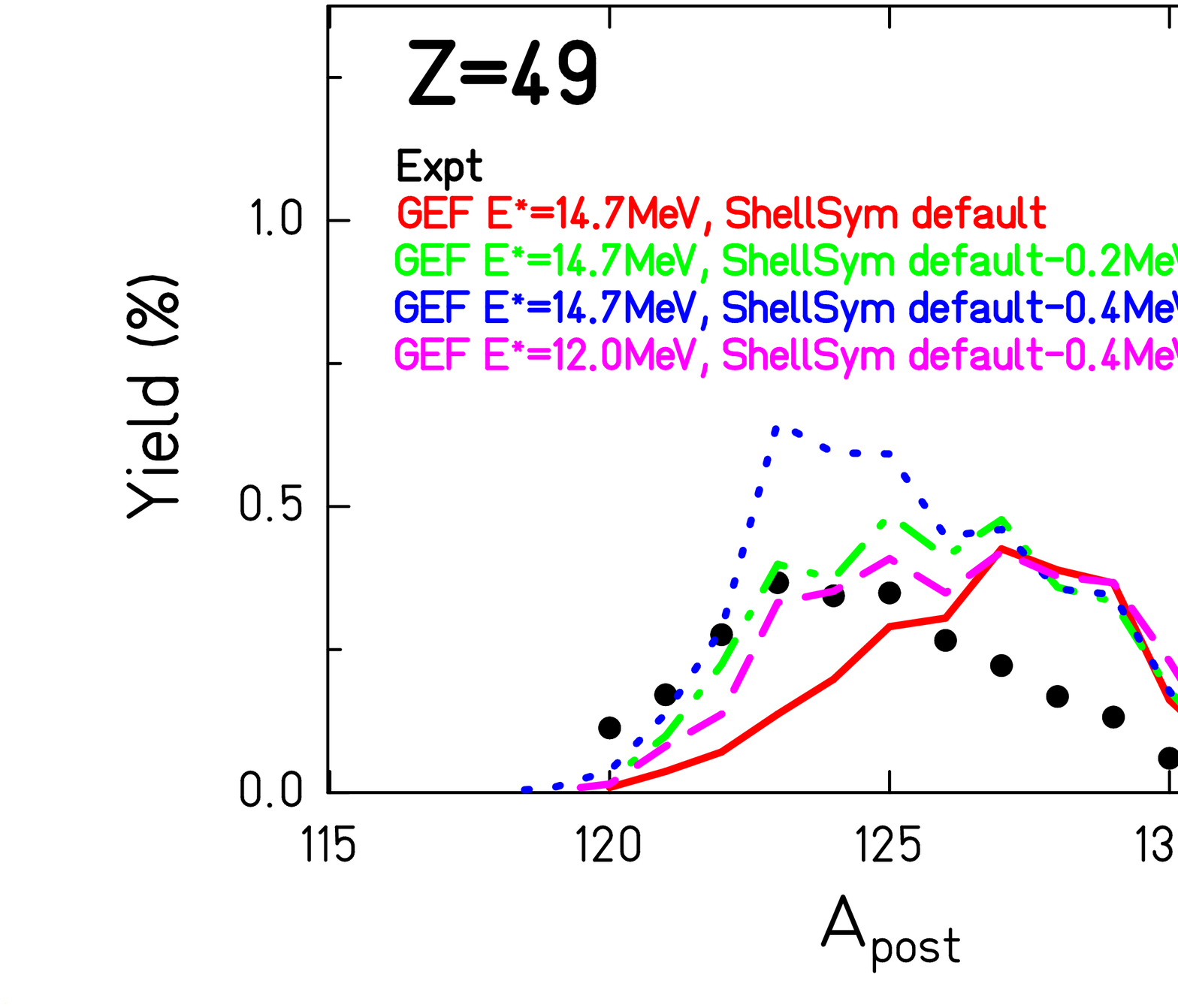}\\
\hspace{-0.8cm}
\includegraphics[width=9.0cm, height=7.cm,angle=0]{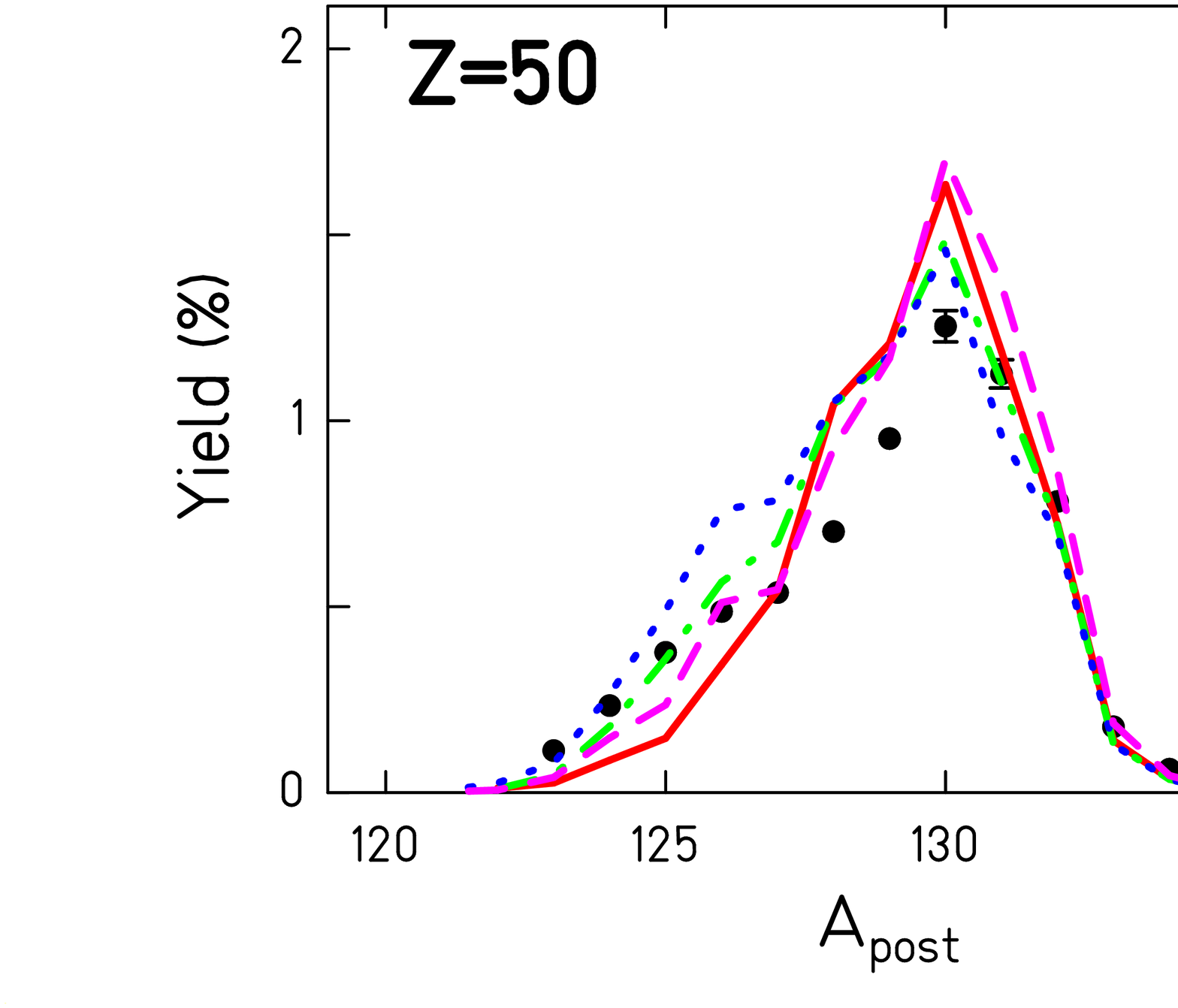}
\vspace{-.30cm}
\caption{(Color online) Fragment post-neutron $A$ distribution for In (top) and Sn (bottom) elements as measured at SOFIA (black dots) in low-energy fission of $^{238}$U \cite{pellereau:2017}. GEF calculations with default parameters (red full line) as well as with enhanced shell correction in the symmetric fission mode (pink dashed, green dashed-dotted and blue dotted lines) are compared. See the text. Whenever not visible, experimental error bars are smaller than the symbols.} 
\label{fig7}
\end{figure}

The experimental approach used at GANIL gives access to fissioning species at and above uranium, and the excitation energy can be selected within a reasonably narrow window \cite{ramos:2016}. However, one fragment is measured, only. In addition, due to close-to-Coulomb barrier beam energy, charge resolution is particularly challenging, and it was observed to be slightly deteriorated at the edge of the distribution. The SOFIA campaign, runned about concomitantly at GSI \cite{boutoux:2013, pellereau:2017}, used collisions between a relativistic primary $^{238}$U, or a secondary actinide, beam and a heavy target to investigate low-energy fission induced by electromagnetic interaction. The set-up permitted to detect both fragments in coincidence, and determine their charges, post-neutron masses and kinetic energies. The SOFIA experiment is complementary to the GANIL one, since the available fissioning systems are located at and below uranium. While resolution is close to perfect at SOFIA \cite{boutoux:2013, pellereau:2017}, the excitation-energy distribution of the fissioning nucleus as excited by electromagnetic interaction consists in quite a wide spectrum from about 5 to 30 MeV \cite{pellereau:2017}, which could not be sorted in the experiment. For the case of the $^{238}$U fissioning system which we consider in this work, model calculations yield a mean excitation energy around 14.7 MeV. Hence, all GEF predictions reported here refer to the fissioning specie $^{238}$U at $E^*$ = 14.7 MeV.\\
The fragment post-neutron neutron number $N$ distribution measured at SOFIA is shown in Fig.~\ref{fig5}. The outcome of the calculation (red full line) describes the measurement reasonably well, including the main structures and the local even-odd staggering. However, the yield around symmetry is observed to be underestimated, consistent with Fig.~\ref{fig1}. This deficiency is further  discussed below. The very different shape of the light and heavy fragment peaks is due to neutron evaporation by the primary fragments after scission. All along the de-excitation cascade, the hot fragments may encounter a magic neutron number. The probability for emitting the next neutron  is considerably reduced whenever such a shell is reached in the cascade. In other words, the fragment $N$ is "trapped" at the shell-stabilized configuration. This is namely the case for the heavy fragments populated in actinide fission which have $N$ close to the $N$ = 82 gap. That explains the intense peak observed at  $N$ = 82, and the relatively narrow width of the heavy-fragment group as compared to the light one. The interpretation is supported by looking at the calculated pre-neutron distribution: Its shape is very similar for the light and heavy fragment groups, overall Gaussian-like and rather "fat". Note that, though, that the pre-neutron mass distribution is {\it not} exactly symmetric about half the mass of the fissioning system, due to multi-chance fission. The fact that the pre- and post-neutron distributions differ much more for the heavy group than for the light one demonstrates the critical influence of the  $N$ = 82 gap in the decay cascade of the heavy fragment.  We note that the staggering observed in the present GEF calculation for the pre-neutron $N$ is reduced as compared to the GEF calculation by Pellereau et al. \cite{pellereau:2017}, namely for the heavy product. This is explained by the different $E^*$ distribution used in the two calculations. As stated above, the present calculation assumes a well-defined $E^*$ = 14.7 MeV. The calculation presented in Ref.~\cite{pellereau:2017} uses as input the wide $E^*$ distribution predicted to be populated in the electromagnetic interaction by a suited model. While the average of that distribution is 14.7 MeV, lower as well as higher $E^*$ enter into play. The lower excitation energy contribution is characterized by larger staggering. On the higher excitation energy side (which extends up to 30 MeV, see Fig.~1c of Ref.~\cite{pellereau:2017}), multi-chance fission plays an increasing role. Evaporation before scission reduces the excitation energy of the finally-fissioning daughter nucleus, implying the revival of strong staggering. In other words, both the energies below and above $<E^*>$ populated in the collision give more structure.

\begin{figure*}[!hbt]
\hspace{-1.cm}
\includegraphics[width=24.cm, height=11.cm,angle=90]{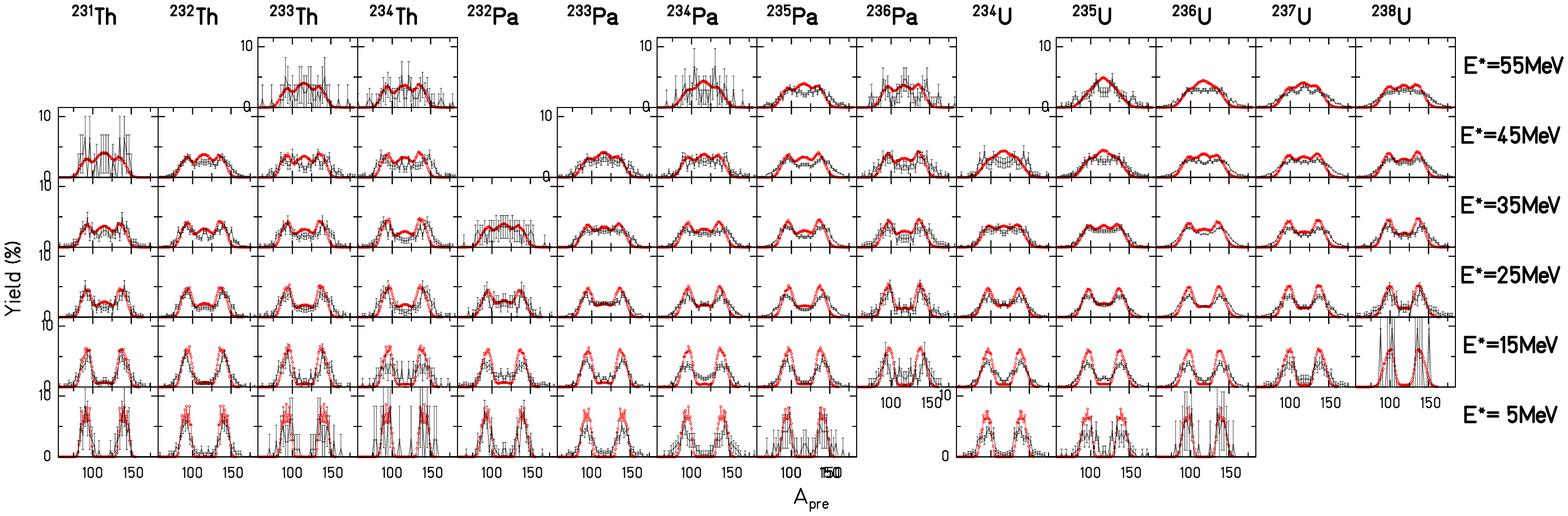}
\vspace{-.10cm}
\caption{(Color online) Fission-fragment pre-neutron $A$ distribution for various systems as measured in Tokai \cite{leguillon:2016} (black crosses) and compared with GEF calculations (red full lines). Fissionioning isotopes are ordered by columns, and mean $E^*$ values are ordered by rows.} 
\label{fig8}
\end{figure*}

Isotopic information from SOFIA is considered in  Fig.~\ref{fig6}. The experimental and calculated post-neutron $N$ distributions for elements between Ge and Pr are shown. The predictions by GEF are confirmed by the experiment for all elements belonging to asymmetric partitions, from Ge to Tc in the light-fragment group, and from Sb to Pr in heavy-fragment one. Discrepancies are obvious around symmetric splitting. An accurate description of the symmetric yield is among most challenging in low-energy fission of actinides, for models of various types (see Refs.~\cite{goutte:2005, randrup:2011, aritomo:2013, pasca:2016} and therein). Due to the very low yield at symmetry, small changes in the prediction can have sizeable impact. In contrast, small changes in predictions for asymmetric partitioning has much less impact, since asymmteric fission is dominating anyhow. Low yields make measurements in the symmetric region challenging as well; and, as already noted in Sec.~\ref{zoller}, discrepancies exist. Uncertainty in experimental information does not light the way for theory. Another difficulty comes from the fact that the symmetric region has also some contribution from the tail of asymmetric fission. Finally, the weight of symmetric fission varies fast with excitation energy, while the asymmetric yields have a milder dependence on $E^*$ \cite{edep-mode}. All together makes a robust prediction of the symmetric yield quite delicate, and it has still to be fixed with suitable data. Work in this direction within the GEF model is in progress. Some illustration is given below.\\
Apart from the overall too low symmetric yield, Fig.~\ref{fig6} shows another very interesting observation: The shape of the experimental isotopic distribution sizeably changes between $Z$ = 49 and $Z$ = 50, from right- becoming left-skewed. In addition, the distribution for $Z$ = 50 - although it is peaked at about $N_{post}$ = 80, exhibits a clear shoulder at lower neutron number around $N_{post}$ = 76. This particular shape is due to the contributions of two fission modes at $Z$ = 50, namely the symmetric mode (shoulder at lower mass) and the so-called S1 asymmetric mode (main peak at higher mass) \cite{pellereau:2017}. Similarly, the $Z$ = 49 distribution can be decomposed into the sum of symmetric and asymmetric components, with the former dominating in this case \cite{pellereau:2017}. The change-over in the skewness of the distribution between $Z$ = 49 and $Z$ = 50 signs the change from dominantly-symmetric to dominantly-asymmetric (S1) fission. Figure~\ref{fig6} interestingly shows that the GEF calculation predicts that the switch from dominantly-symmetric to dominantly-aysmmetric splitting occurs already between $Z$ = 48 and $Z$ = 49. That shows the limitation of the model in accurately describing the competition between symmetric and asymmetric fission, already pointed out above. A discussion on this point, and what it can learn us about the underlying physics is proposed below.\\
We start with reminding that shell effects are expected also at symmetry (at low excitation energy) \cite{mulgin:1998, schmidt:2000}.  In contrast to the asymmetric fission modes (so-called S1 and S2) which are established to be well localized in $Z$, independent of the fissioning system \cite{bockstiegel:2008} due to strong shell effects, the locus of symmetric fission depends on the size (neutron and proton numbers) of the fissioning nucleus. Depending on that size, symmetric fission is, or is not, itself located in a region of shell-stabilized fragments. That means that it can be either strengthened or attenuated by some weak shells. In the GEF code, this possibility is accounted for by a parameter "Shell effect in the symmetric channel" (hereafter, ShellSym). This parameter shall in principle vary from system to system. It is not easy to determine from so-far available experimental information: In many actinides the yields at symmetry for thermal neutron energies are very low and not well measured. Data with fast neutrons are not suited, because they often have broad energy distributions. The yield of the symmetric channel has thus quite some uncertainty, if there are no suitable data to fix the shell effect at symmetry. According to this uncertainty, it was decided to assume in GEF a default value of ShellSym = +0.3 MeV (weak anti-shell), and the parameter is assumed the same for all systems in the absence of robust empirical information. However, this value can certainly vary by a few 100 keV depending on the system, as discussed above. Its influence is illustrated in Fig.~\ref{fig7}, where the experimental post-neutron mass distributions for $Z$ = 49 and $Z$ = 50 are compared with various GEF calculations: $E^*$ = 14.7 MeV with ShellSym=+0.3 MeV (default like in Fig.~\ref{fig6} - red full line),  ShellSym= -0.2 MeV (green dashed-dotted line), and ShellSym= -0.4 MeV (blue dotted line). We display also a calculation with $E^*$ = 12.0 MeV for ShellSym=-0.4 MeV (pink dashed line) ; this is to show the additional influence of excitation energy within the experimental interval. When modifying the ShellSym parameter, and namely creating a shell-stabilized effect for the symmetric mode (ShellSym negative), the isotopic distribution of $Z$ = 49 becomes left-skewed, and a shoulder appears in the $Z$ = 50 distribution, in accordance with experimental information. Perfect quantitavive description of the measurement is not reached yet. A possible explanation is the $E^*$ spread which has to be accounted for simultaneously, namely for the yield of the different fission modes (symmetric, S1 and S2). This is beyond the goal of the present work.\\
We emphasize that these results are absolutely new. It is the first time that complete, high-resolution $Z$ information is reached near symmetry, over the entire (light and heavy fragment) production at these excitation energies. Combining the new and unique data available now with SOFIA, and a model like GEF - which is in agreement with previously available data of sometimes limited choice and accuracy, the transition region in the fragment distribution from the asymmetric peaks to the symmetric mode channel, and concomitantly the evolution of shell effects, can be studied. Work along this line is in progress, and is expected to help improving further the model.\\
Besides fragment isotopic yields, GEF calculations were compared with SOFIA data on TKE and neutron multiplicities \cite{martin:2015}. A high degree of correlation between these observables and with fragment isotopic yields is seen in the data and in the calculation \cite{schmitt:2017}. A model like GEF, which preserves the link betwen fission quantities, is suited to study the physics mechanism behind the experimental observation as it gives access to observables that are not (yet) accessible in the measurement.

\begin{figure*}[h]
\hspace{-1.8cm}
\includegraphics[width=7cm,height=14.2cm]{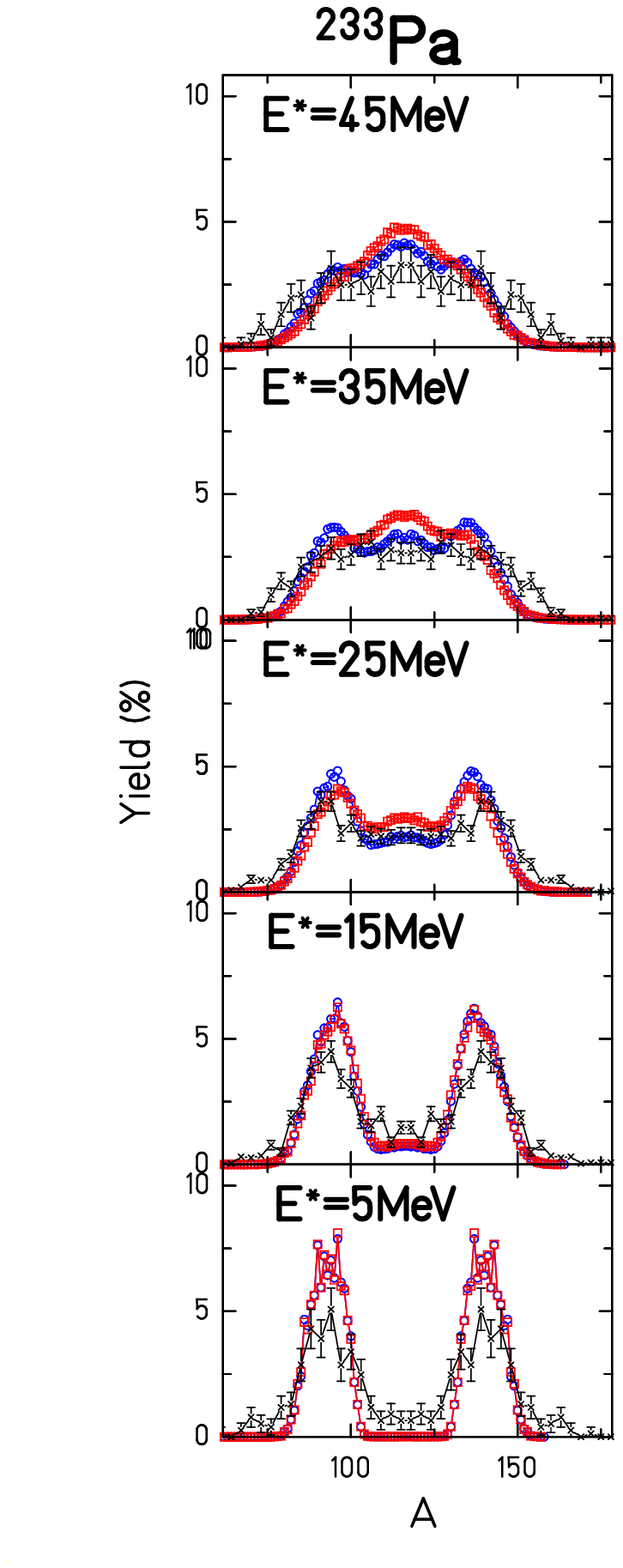}
\hspace{-1.9cm}
\includegraphics[width=7cm,height=14.2cm]{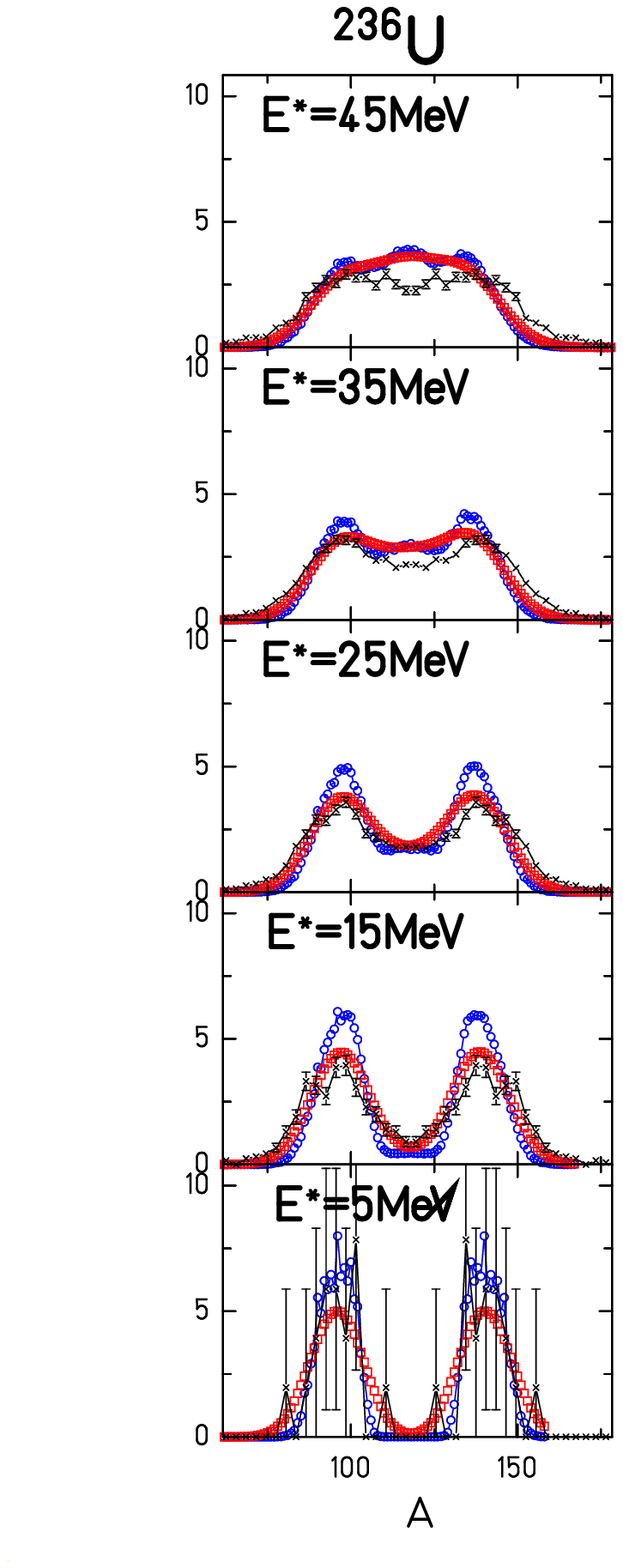}
\hspace{-1.9cm}
\includegraphics[width=7cm,height=14.2cm]{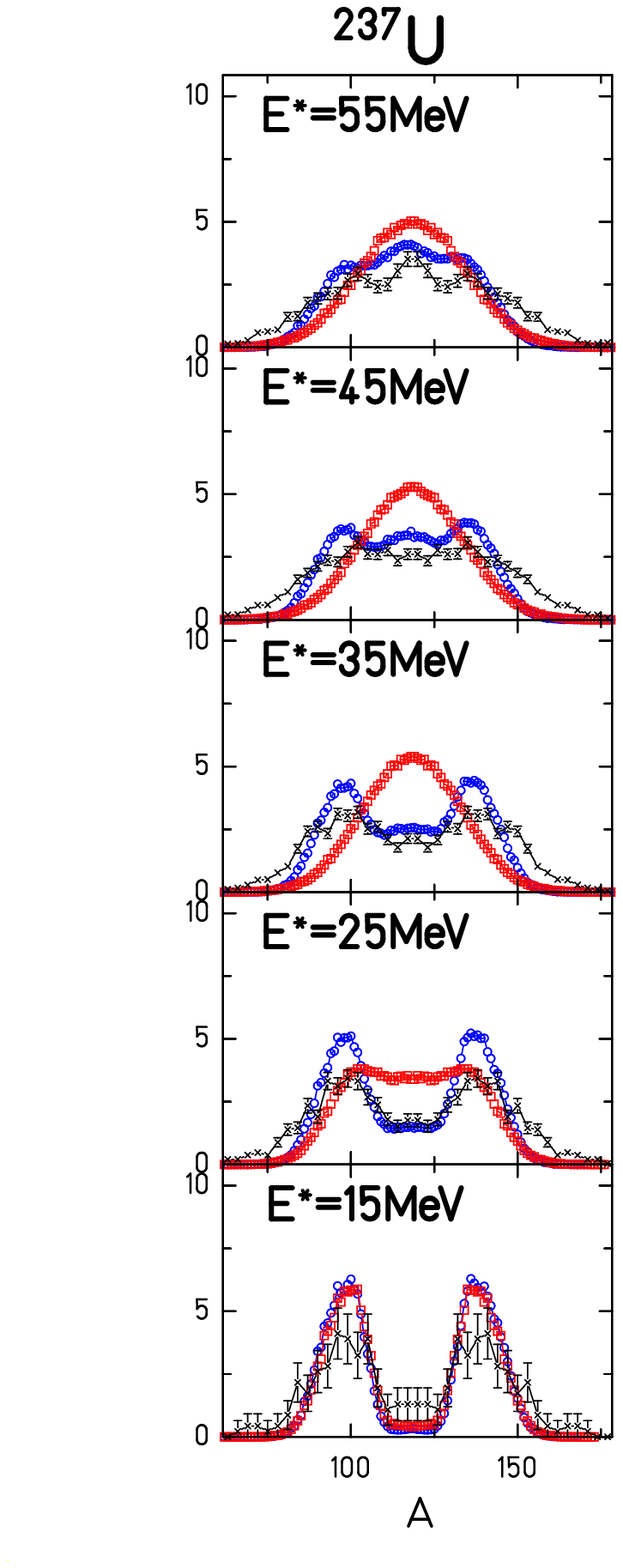}
\vspace{-0.2cm}
\caption{Fission-fragment pre-neutron $A$ distribution for selected systems from Ref.~\cite{leguillon:2016} (black crosses) compared with GEF calculations. First column: calculations with $L$=0 (blue diamonds) and $L$=20 (red squares) as a typical maximum value in transfer; second column: calculations assuming unique mass resolution (blue diamonds) and folded with the experimental resolution of $\sigma$ = 6 units (red squares); third column: calculations without (red circles) and with (blue diamonds) multi-chance fission taken into account.}
\label{fig9}
\end{figure*}

\subsection{Multi-nucleon transfer-induced fission of Th, Pa, and U with $E^*$ from threshold to 60 MeV}\label{nishio}

\begin{figure}[h]
\hspace{-1.40cm}
\includegraphics[width=9.cm,height=7.cm]{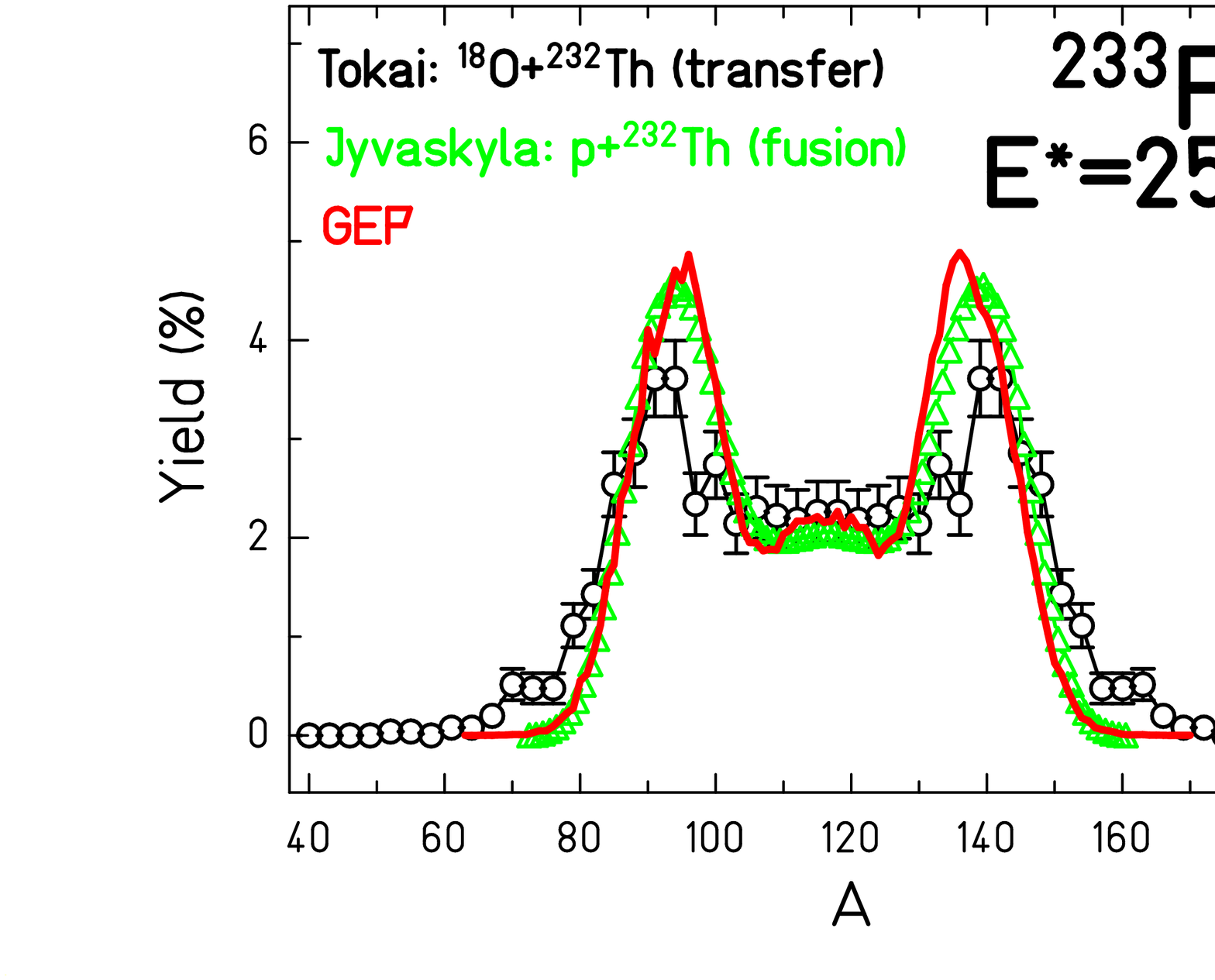}
\vspace{-0.5cm}
\caption{Fission-fragment pre-neutron $A$ distribution for $^{233}$Pa at $E^*$ = 25 MeV as measured at 
Tokai \cite{leguillon:2016} (black circles) and at Jyvaskyla \cite{kozulin:2014} (green triangles) in transfer- and proton-induced fission, respectively. The calculation with GEF (red full line) for $L$ = 0 and $E^*$ = 25 MeV is shown also.}
\label{fig10}
\end{figure}

An ongoing experimental campaign at JAEA Tokai \cite{leguillon:2016, nishio:2017} is being dedicated to the systematic study of fission-fragment mass distributions over a very wide range of systems by exploiting multi-nucleon transfer. The results from $^{18}$O+$^{232}$Th collisions were published in Ref.~\cite{leguillon:2016}. Leguillon et al.  measured in a single experiment the pre-neutron mass distribution for about 15 fissioning isotopes between $^{231}$Th and  $^{238}$U, over a wide $E^*$ range, from threshold to about 60 MeV, and angular momentum $L$. Beside its size, such a data set is also precious because all systems are measured and analyzed under the same conditions, ensuring consistency between them (avoiding the sometimes difficult comparison of different approaches). As compared to the experiment at VAMOS, both fragments are detected at Tokai, but only masses are identified, and with limited resolution. On the other hand, the domain covered in fissioning system ($A$, $Z$, $E^*$, $L$) is considerably extended. That permits to investigate the performance of GEF in a region which could not be probed so far, and in a systematic way.\\
The comparison of the data set of Ref.~\cite{leguillon:2016} and model calculations is displayed in Fig.~\ref{fig8}. Note that, in the measurement, the initial excitation energy was sorted into bins which are 10 MeV-wide about the mean value quoted in the figure. The calculation was done at fixed $E^*$ taken equal to this mean value. The survey of the chart shows that GEF describes the evolution of the distribution with the $N$ (equivalently, $A$) and $Z$, as well as the $E^*$ of the initial system. Detailed inspection, though, shows some deviations. These may be attributed to several reasons \cite{schmitt:2017}: (i) the angular momentum imparted to the fissioning nucleus which depends on the transfer channel (in the calculation of Fig.~\ref{fig8}, $L$ was set to zero), (ii) the spread in $E^*$ inherent to the experiment, where (the calculation assumes a well-defined $E^*$, (iii) the limited mass resolution ($\sigma$ = 6 units quoted in Ref.~\cite{leguillon:2016}). Finally, model deficiencies are not excluded neither.\\
To estimate the quantitative influence of some of the aforementioned effects, the GEF calculations were repeated at various $L$ values. Also, they were folded with the experimental resolution. Typical results are shown in Fig.~\ref{fig9}.  The model suggests that the influence of angular momentum (first column), although it is present, cannot be clearly resolved within the experimental error bars for the present systems \footnote{We note that some influence of $L$ could have been discussed also in the context of the VAMOS experiment. Yet, since the influence is observed to be small for collisions between actinides and oxygen, it is expected to be even weaker for actinides on carbon, and was therefore not mentioned in Sec.~\ref{vamos}.}. A similar observation was made regarding the influence of the finite width of the $E^*$ window \cite{schmitt:2017}. Experimental mass resolution is seen to have the dominant influence (second column). Figure~\ref{fig9} also illustrates the  crucial need to account for multi-chance fission (third column), at excitation energies as low as about 20 MeV. That is consistent with the finding based on GEF calculations in Ref.~\cite{schmidt:2016}, as well as with work based on different theoretical approaches \cite{nishio:2017, nishio:2017b, aritomo:2013, moller:2017}.\\
The expected critical role of mass resolution is further investigated in Fig.~\ref{fig10}. The pre-neutron mass distribution obtained at Tokai (black circles) for fission of $^{233}$Pa at $E^*$ = 25 MeV is overlaid with a measurement performed at Jyvaskyla (green triangles) \cite{kozulin:2014} for the same system. In the latter experiment, the fissioning nucleus was produced by fusion of a proton with $^{232}$Th. This mechanism implies a well-defined initial excitation energy and a low angular momentum. The set-up of Ref.~\cite{kozulin:2014} permitted to achieve a mass resolution of $\sigma$ around 1.5 unit, and statistics was high. The comparison of Fig.~\ref{fig10} shows clear deviation between the two data sets. Good agreement between the Jyvaskyla measurement and the GEF calculation\footnote{The calculation shown is not folded with the experimental resolution $\sigma$ = 1.5. Doing so yields a curve which is hardly distinguishable on the scale of the figure.} (red full line) is observed. Combined with Fig.~\ref{fig9}, this agreement suggests, based on experimental facts, a dominant contribution of resolution, and which hinders the study of "finer" angular momentum and excitation energy effects from the data collection of Leguillon et al. \cite{leguillon:2016} alone. At the same time, as noted previously, the strategy pursued at Tokai gives access in reasonable measurement times to a wide range of fissioning system mass, charge and excitation energy, where other approaches like {\it e.g.} Ref.~\cite{kozulin:2014} can study only one system at a time.\\   
The above discussion further suggests that the GEF code is a relevant tool to trace back the different influences - of either physical or experimental origin, and possibly unfold each of them in the measurement. New data from Tokai \cite{nishio:2017} from various multi-nucleon transfer reaction are becoming available. Cross bombardment, {\it i.e.} the fomation of a specific compound nucleus by means of different entrance channels, allows to vary independently $E^*$ and $L$ for a given fissioning system ($A$, $Z$), and {\it vice and versa}. The potential of GEF can be exploited further in this context to discriminate between the various influences, provided resolution and statistics are sufficient. That is important to avoid mis-interpretation.

\subsection{Specific fragment yields after radioactive decay}\label{decay}

Nuclear applications strongly rely on the values of some specific fission yields, as produced after long times, {\it i.e.} after potential radioactive decay of the fragments formed by the reaction. These yields, often called fission product yields (FPY), are crucial at nuclear reactors due to potential radiotoxicity. In this context, decay heat and emission of delayed neutrons have to be controlled as well. As for nuclear medicine, very specific radionuclides can be used as diagnostic, when not to destroy malignant tumors. The fission yields calculated with GEF can feed dedicated codes which were developed for applications to predict FPY, decay heat, delayed neutrons and $\gamma$-rays (see Ref.~\cite{schmidt:2016} and therein). 

\begin{figure}[h]
\hspace{-1.40cm}
\includegraphics[width=9.5cm,height=12.cm]{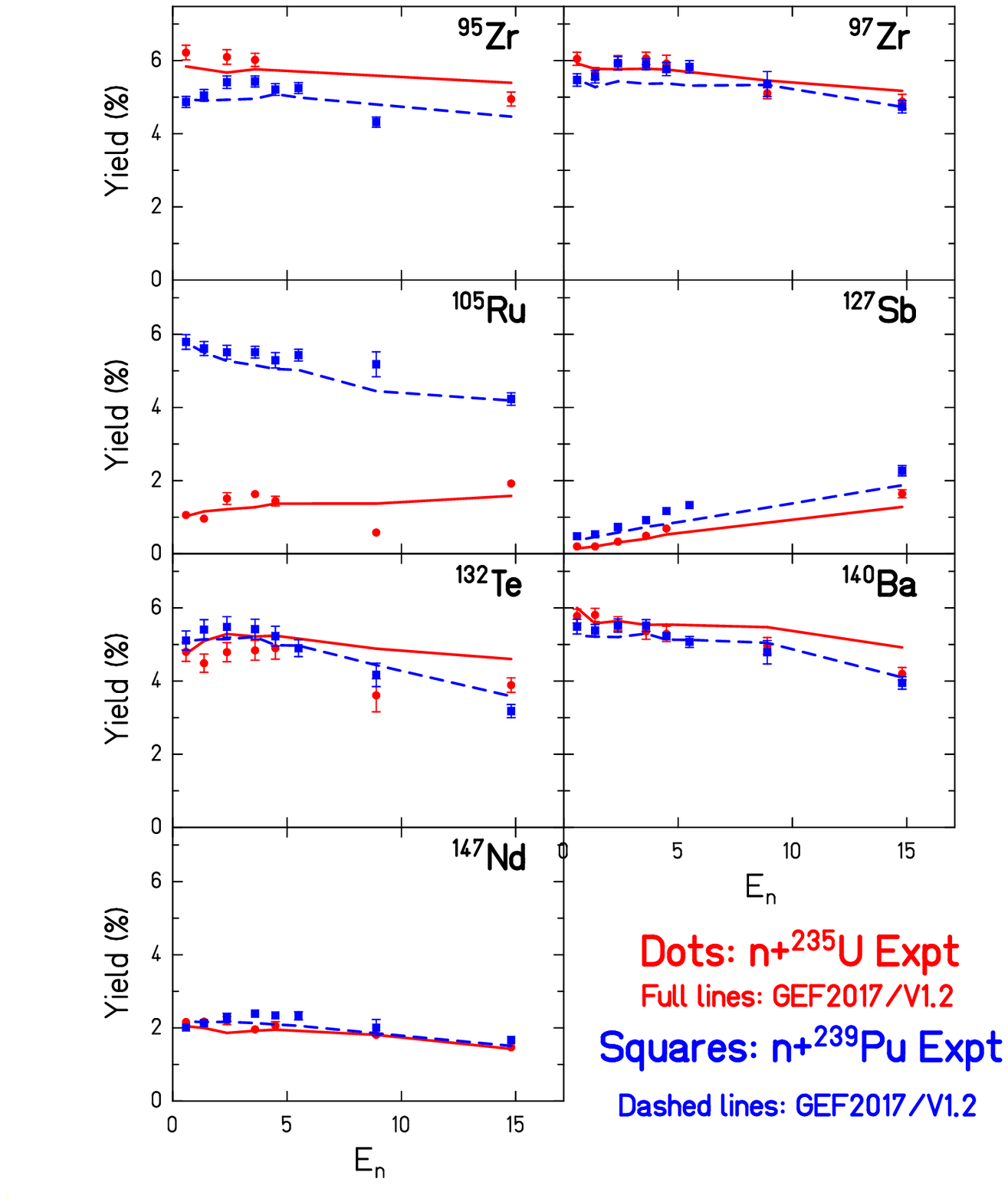}
\vspace{-0.5cm}
\caption{Experimental (symbols) and calculated (lines) FPY for specific radionuclides as a function of incident energy for neutron-induced fission of $^{235}$U (red dots) and $^{239}$Pu (blue squares). Experiment is taken from Ref.~\cite{gooden:2016}.}
\label{fig11}
\end{figure}

As an example of the utility of the GEF in the aforementioned fields, Fig.~\ref{fig11} displays the fission product yield for several radioisotopes as a function of incident energy for neutron-induced fission of $^{235}$U and $^{239}$Pu \cite{gooden:2016}. Comparison with GEF calculations shows overall satisfactory agreement. We note that, provided that the available empirical information on the decay properties of a specific product is sufficiently known, the description mostly relies on the capability of GEF to produce fission yields that are close to reality.

\section{Summary and conclusions}\label{concl}

The achievement of the GEneral description of Fission observables model, GEF, which has proven good for so-far existing data, is extended and studied in detail, in combination with the appearence of new data sets. The fragment isotopic distributions predicted by GEF for various fissioning isotopes and excitation energies are found to well describe recent measurements. Remaining discrepancies are investigated, demonstrating the potential of GEF for improving the knowledge in the transition region between asymmetric and symmetric fission, and the influence of shell effects. Description of more and more exclusive observables is crucial for improving our understanding, due to the high-degree of correlation between the various fission quantities.\\
The evaluation of the model is further extended to a wide region of the nuclear chart, under various so-far un-probed conditions. While the code explains consistently the main trends as function of fissioning system mass, charge, and excitation energy, its predictions are proposed to be used to discriminate between various physical effects, on one side, and experimental bias, on the other side. The study highlights the need of high-quality experimental information for probing fission in finer detail. A model like GEF, with a unique set of parameters based on physical grounds, and which does not rely on any specific empirical input for the particular system to be considered, is a relevant tool to assist experimental analysis and guide data interpretation in this respect. 

{\bf Acknowledgements}
\\
We thank Dr. Diego Ramos and Dr. Hirose Kentaro for providing the experimental points measured at GANIL and JAEA, respectively. The work was supported by the French-German collaboration between IN2P3-DSM/CEA and GSI, under agreement 04-48.

\end{document}